%% file: ms.tex
\documentclass{aa}  
\usepackage[colorlinks=true,urlcolor=blue,citecolor=blue]{hyperref}
\usepackage{graphicx}
\usepackage{txfonts}
\usepackage{fixltx2e}
\usepackage{lscape}
\usepackage{multirow}

\begin{document}

\title{The International Deep Planet Survey II:\\ The frequency of directly imaged\\ giant exoplanets with stellar mass\thanks{Tables 11, 13, 14 and 15 are available in electronic form at the CDS via anonymous ftp to cdsarc.u-strasbg.fr (130.79.128.5) or via \href{http://cdsweb.u-strasbg.fr/cgi-bin/qcat?J/A+A/}{http://cdsweb.u-strasbg.fr/cgi-bin/qcat?J/A+A/}}}

\author{R. Galicher\inst{1,2,3}, C. Marois\inst{3,4}, B. Macintosh\inst{4,5}, B. Zuckerman\inst{6}, T. Barman\inst{7}, Q. Konopacky\inst{8}, I. Song\inst{9}, J. Patience\inst{10}, D. Lafreni\`{e}re\inst{11}, R. Doyon\inst{11}, E. L. Nielsen\inst{5,12}}
\institute{LESIA, Observatoire de Paris, CNRS, Universit\'e Paris Diderot, Universit\'e Pierre et Marie Curie, 5 place Jules Janssen, 92190 Meudon, France\\
\email{raphael.galicher at obspm.fr}
\and
Groupement d’Int\'er\^et Scientifique PHASE (Partenariat Haute r\'esolution Angulaire Sol Espace) between ONERA, Observatoire de Paris, CNRS and Universit\'e Paris Diderot
\and 
National Research Council Canada, 5071 West Saanich Road, Victoria, BC, V9E 2E7, Canada
\and
Lawrence Livermore National Laboratory, 7000 East Ave., Livermore, California, 94550, USA
\and
Kavli Institute for Particle Astrophysics and Cosmology, Stanford University
\and
Department of Physics and Astronomy, University of California, Los Angeles, CA 90095, USA
\and
Lunar and Planetary Laboratory, University of Arizona, Tucson AZ 85721 USA
\and
CASS, University of California San Diego, La Jolla, CA 92093-0424, USA
\and
Department of Physics and Astronomy, University of Georgia, Athens, GA 30602-2451, USA
\and
Arizona State University, Tempe, AZ 85281, USA
\and
D\'epartement de Physique, Universit\'e de Montr\'eal, C.P. 6128 Succ. Centre-Ville, Montr\'eal, QC H3C 3J7, Canada
\and
SETI Institute, Carl Sagan Center, 189 Bernardo Avenue, Mountain View, CA 94043, USA}

   \date\today

\titlerunning{IDPS Survey}
\authorrunning{Galicher et al.}

\abstract
{Radial velocity and transit methods are effective for the study of short orbital period exoplanets but they hardly probe objects at large separations for which direct imaging can be used.}{We carried out the international deep planet survey of 292 young nearby stars to search for giant exoplanets and determine their frequency.}{We developed a pipeline for a uniform processing of all the data that we have recorded with NIRC2/Keck II, NIRI/Gemini North, NICI/Gemini South, and NACO/VLT for 14 years. The pipeline first applies cosmetic corrections and then reduces the speckle intensity to enhance the contrast in the images.}{The main result of the international deep planet survey is the discovery of the HR\,8799 exoplanets. We also detected 59 visual multiple systems including 16 new binary stars and 2 new triple stellar systems, as well as $2,279$ point-like sources. We used Monte Carlo simulations and the Bayesian theorem to determine that $1.05\rlap{\textsuperscript{\tiny+2.80}}\textsubscript{\tiny-0.70}\ \%$ of stars harbor at least one giant planet between $0.5$ and $14$\,M$_\mathrm{J}$ and between $20$ and $300$\,AU. This result is obtained assuming uniform distributions of planet masses and semi-major axes. If we consider power law distributions as measured for close-in planets instead, the derived frequency is $2.30\rlap{\textsuperscript{\tiny+5.95}}\textsubscript{\tiny-1.55}\ \%$, recalling the strong impact of assumptions on Monte Carlo output distributions. We also find no evidence that the derived frequency depends on the mass of the hosting star, whereas it does for close-in planets.}{The international deep planet survey provides a database of confirmed background sources that may be useful for other exoplanet direct imaging surveys. It also puts new constraints on the number of stars with at least one giant planet reducing by a factor of two the frequencies derived by almost all previous works.}

        
    \maketitle
    
\section{Introduction}
After the first discoveries of exoplanets by indirect detections in the late 80s and 90s, several teams performed surveys to obtain direct images of substellar objects in the optical and near-infrared. As the instruments were not optimized for high contrast imaging, the first surveys probed brown dwarfs that are brighter than planets and thus, easier to detect. For example, \citet{becklin88}, \citet{schroeder00}, \citet{gizis01}, \citet{oppenheimer01}, and \citet{carson09} observed nearby stars, while \citet{neuhauser04} and \citet{lowrance05} surveyed young systems, and \citet{chauvin06} targeted stars harboring planets detected by stellar radial velocity.
Taking advantage of new observing modes, such as spectral differential imaging \citep[SDI;][]{marois00} and angular differential imaging \citep[ADI;][]{marois06} and using more powerful adaptive optics systems, later surveys of youthful stars probed fainter and fainter substellar companions down to the young gas-giant extrasolar planet regime \citep{biller07,kasper07,lafreniere07a,nielsen08,metchev09,chauvin10,heinze10,rameau13a,yamamoto13,biller13,nielsen13,wahhaj13,brandt14}.

For 14 years, we had been using Keck II, Gemini North and South, and VLT to run two near-infrared imaging surveys: a Keck adaptive optics search for young exoplanets \citep{kaisler03} and the international deep planet survey \citep[IDPS;][]{marois10c}. In this paper, we merge the two and call IDPS the resulting survey. The targets were 292 young and nearby stars of A to M spectral types with a majority of massive stars. The main objectives of the IDPS were the detection and spectral characterization of new exoplanets, and the determination of the frequency of stars harboring giant planets with long orbital periods.

A first paper \citep{vigan12} presented a fraction of the A stars of the IDPS. In the current paper, we present the complete survey.  The target sample is described in \S\,\ref{sec:target}. The observations and instruments that we used are presented in \S\,\ref{sec:obs}. Then, \S\,\ref{sec:pipeline} details the pipeline that we developed for the uniform data processing of the $\sim$30,000 frames. The characteristics of all the detected sources (multistellar systems, exoplanets, and field stars) are listed in \S\,\ref{sec:results}. Finally, we run a Monte Carlo analysis in \S\,\ref{sec: plafreq} to constrain the frequency of stars with giant planets.

\section{Target sample}
\label{sec:target}
\begin{figure}[!ht]
\centering
\includegraphics[width=0.4\textwidth]{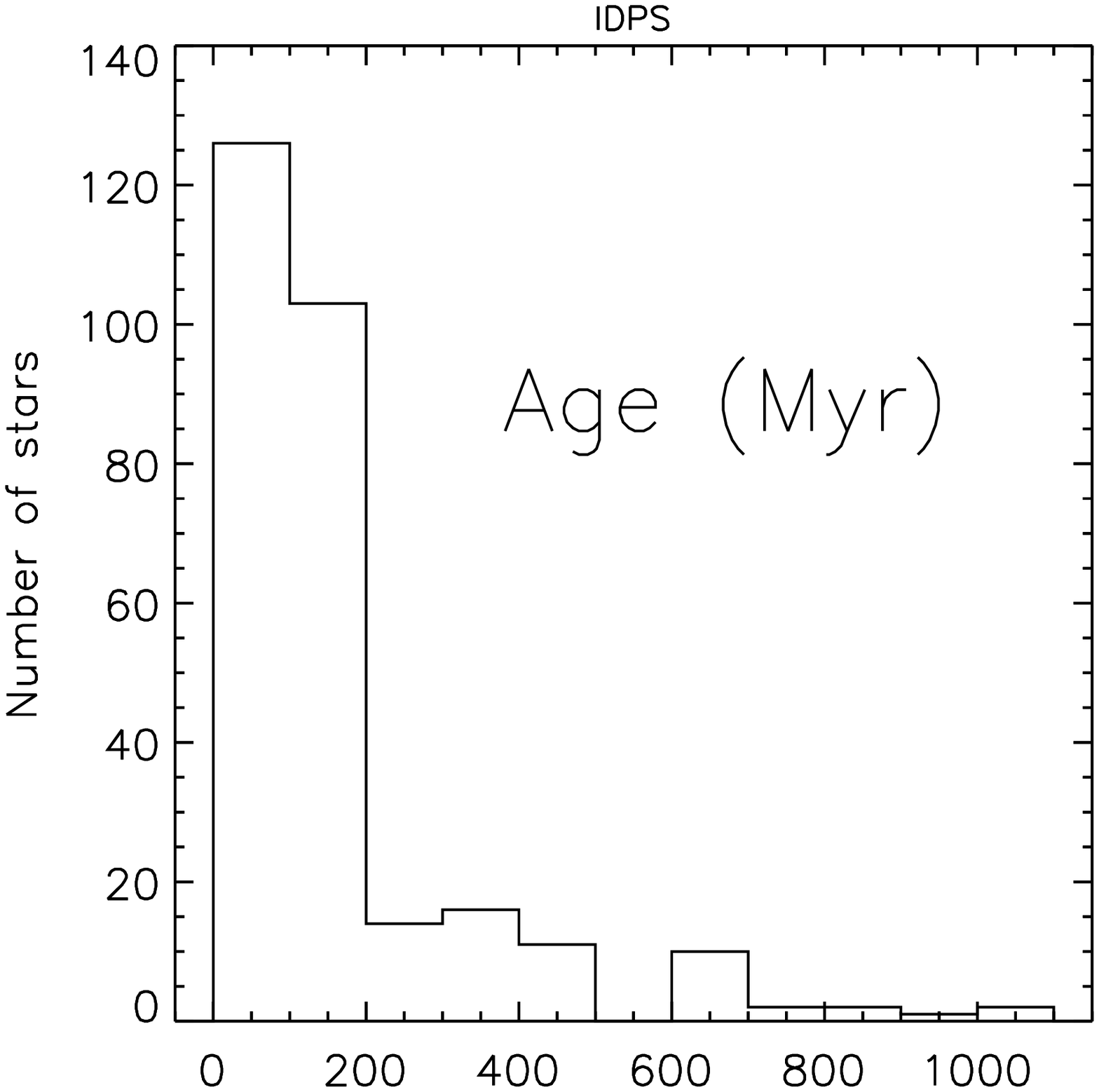}\\
\includegraphics[width=0.4\textwidth]{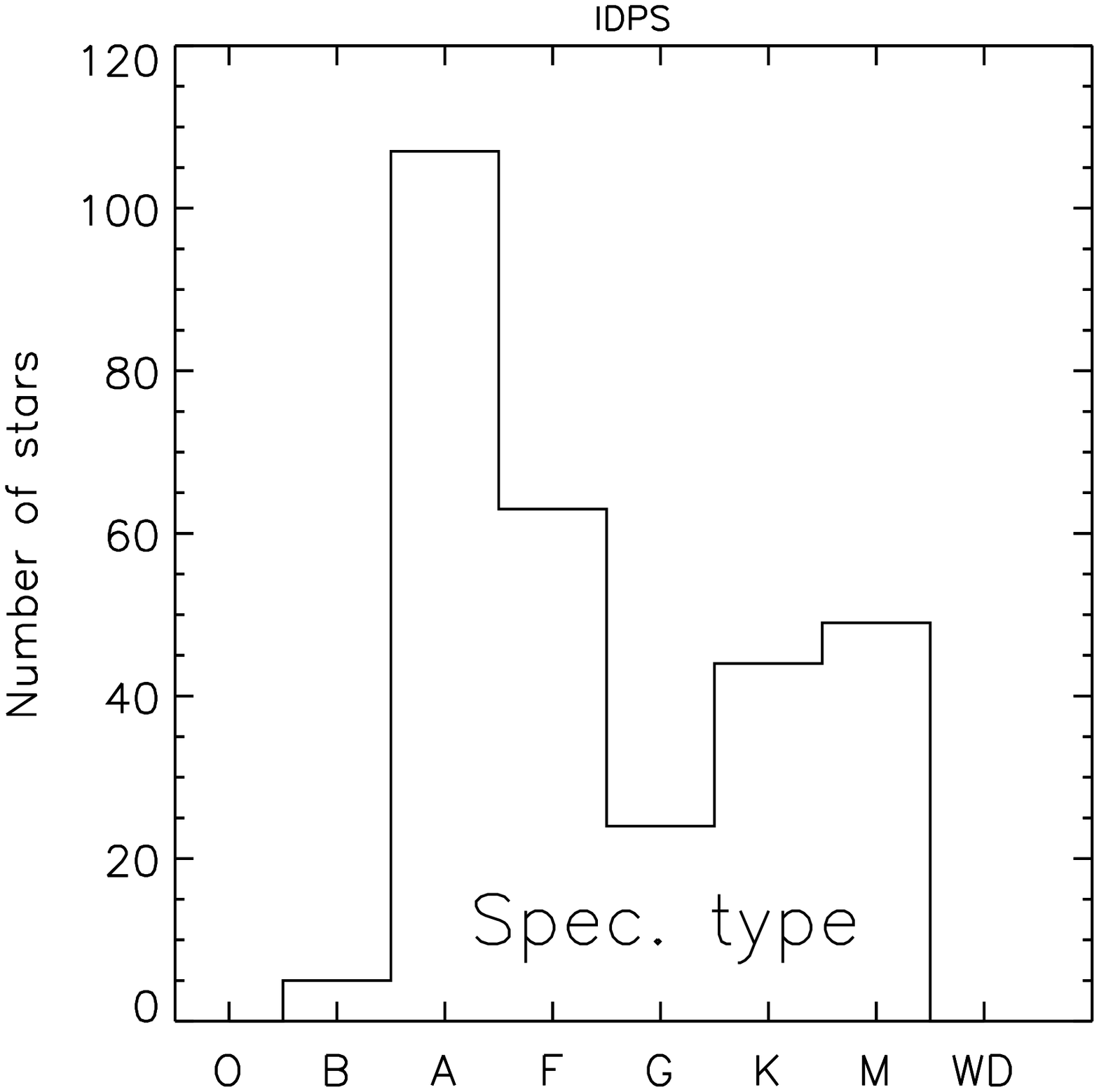}\\
\includegraphics[width=0.4\textwidth]{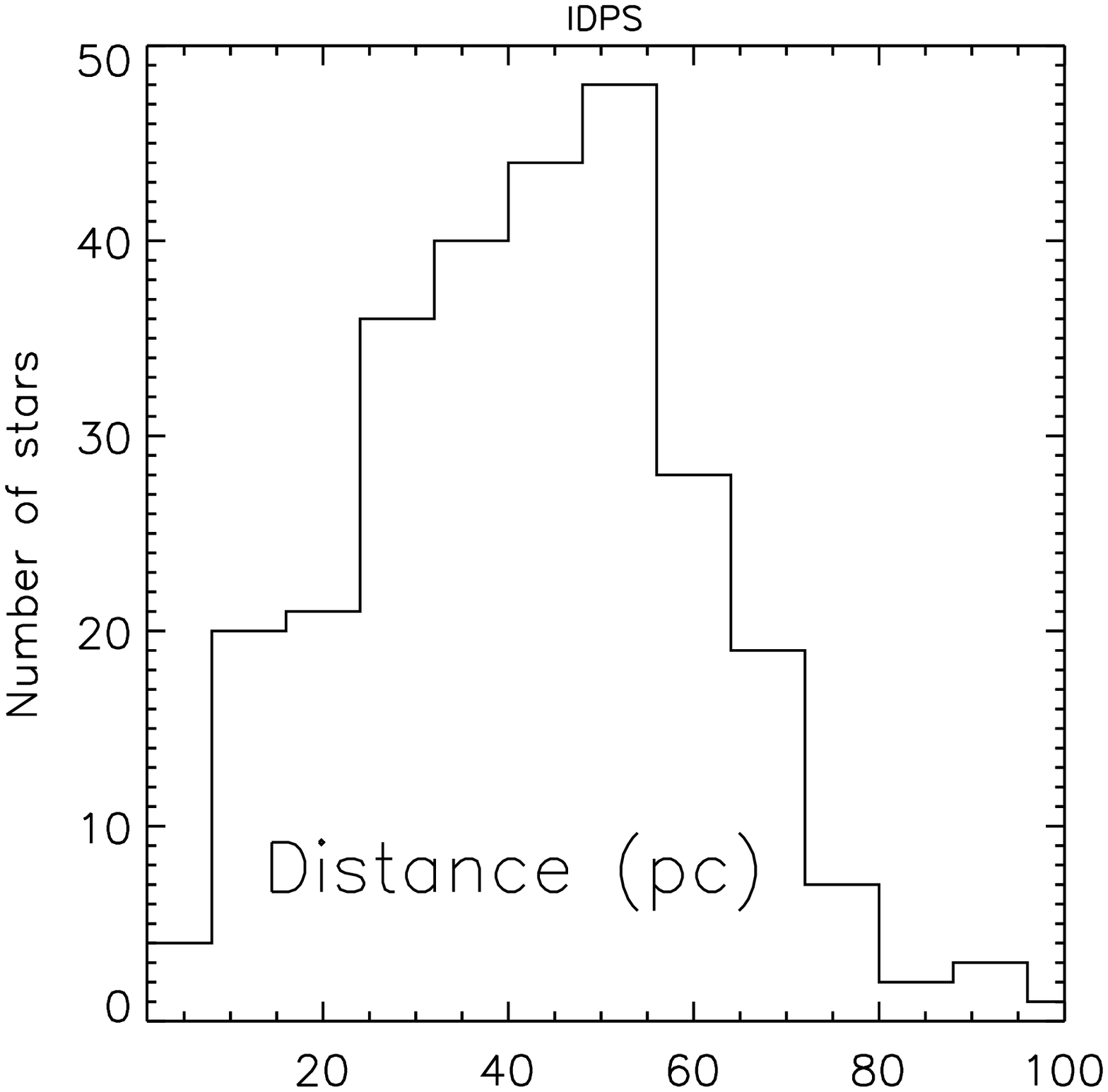}
\caption[]{\sl IDPS star age, spectral type, and distance distributions.}
\label{fig: starsample}
\end{figure}

As for most of exoplanet direct imaging surveys, the target sample is composed of young and nearby stars (Tab.\,\ref{tab: targlist0}) with a median age of $120$\,Myr and a median distance of $45$\,pc (Fig.\,\ref{fig: starsample}).
\begin{table*}
\begin{center}
\begin{tabular}{lllcccccccc}
Star&$\alpha$&$\delta$&Spec.&Dist.&\multicolumn{3}{c}{Age (Myr)}&Age method&Ref&IR\\
&(J2000.0)&(J2000.0)&type&(pc)&median&min&max&or Assoc.&&excess\\
\hline
HIP682&00:08:25.75&+06:37:00.49&G2V&40&100&30&300&Li/X&&Y\textsuperscript{c}\\
QTAND&00:41:17.34&+34:25:16.87&Ge&41&100&30&300&Li/X/UV&&--\\
\hline
\end{tabular}
\caption{\sl Target list. The complete list is given in Tab.\,\ref{tab: targlist}.}
\label{tab: targlist0}
\end{center}
\end{table*}
The youth ensures that giant planets are bright enough in the near-infrared to be detected assuming evolutionary models like COND and DUSTY \citep{chabrier00,baraffe03}. The ages in Tab.\,\ref{tab: targlist0} are extracted from previous work (Ref. column) or were derived by our team using the following techniques described in \citet{zuckerman04}: lithium, X-ray, UV emission, UVW velocity, H-alpha, and color-magnitude diagrams.

All types of main-sequence stars were surveyed with a majority of massive stars: 5, 107, 63, 24, 44, and 49 B, A, F, G, K, M stars. The complete sample is composed of 292 stars.

\section{Observations}
\label{sec:obs}
We obtained near-infrared observations with the adaptive optics systems of Gemini North, Gemini South, Keck II, and VLT. We used public HST data to confirm one of our candidates to be a background object. The observations are described in the following sections and in Tab.\,\ref{tab: obslist0}. The information about the NACO/VLT observations are gathered in \citet{vigan12} and they are not included in this paper although we use the associated detection limits for the statistical study (\S\,\ref{sec: plafreq}).
\begin{table*}
\begin{center}
\begin{tabular}{lcccccccc}
Star&Date&Nb$\times$Exp$\times$Coadd&FOV rot&Tel&Filter&Config&Mode&Program\\
&&(s)&(deg)&&&Coro&&\\
\hline
HR9&2001-12-01&8$\times$15$\times$4&4.4&KE&Kp&w&PA&\\
TYC1186-706-1&2008-12-17&9$\times$0.5$\times$10&1.0&KE&Kcont&-&PA&A162N2\\
HIP107350&2005-07-18&6$\times$30$\times$1&10.2&GN&CH4s&LO&V&GN-2005A-Q-16\\
HIP107412&2007-09-12&139$\times$30$\times$1&55.6&GN&CH4s&LI&V&-\\
\hline
\end{tabular}
\caption{\sl Observation list. The complete list is given in Tab.\,\ref{tab: obslist}.}
\label{tab: obslist0}
\end{center}
\end{table*}

\subsection{NIRI at Gemini North}
At Gemini North, we observed with the NIRI camera~\citep{hodapp03} and the ALTAIR adaptive optics system~\citep{herriot00}. Data were taken from 2004 to 2014~(programs GN-2007B-Q-59, GN-2008A-Q-77, GN-2008B-Q-64, GN-2009A-Q-80, GN-2009B-Q-17, GN-2011B-Q-11, GN-2012B-Q-14, GN-2012B-Q-70, GN-2013A-Q-34, GN-2013B-Q-16, and GN-2014A-Q-18).

The f/32 camera was used, yielding a $22"\times22"$ field of view and 1024$\times$1024 pixel images. The spatial sampling is 21.4\,mas or 22.0\,mas per pixel if the Altair field lens is used~(LI in Tab.\,\ref{tab: obslist0}) or not~(LO). The lens improves the off-axis AO correction but induces astrometric distortions (Sec.\,\ref{subsec: firststep}). Table\,\ref{tab: filt} gives the specifications of the spectral filters that were used.
\begin{table}[!ht]
\begin{center}
\begin{tabular}{cccc}
\hline
Tel.&Name&$\lambda_0$&$\lambda_\mathrm{min}-\lambda_\mathrm{max}$\\
&&$(\mu$m)&or $\Delta\lambda$\\
\hline
\multirow{6}*{GN}&J&1.25&1.15-1.33\\
&H&1.65&1.49-1.78\\
&CH4s\textsuperscript{1}&1.58&6.5\%\\
&K&2.20&2.03-2.36\\
&Ks&2.15&1.99-2.30\\
&Kp&2.12&1.95-2.30\\
\hline
\multirow{2}*{GS}&CH4-H4\%S&1.578&4.00\%\\
&CH4-H4\%L&1.652&3.95\%\\
&Ks&2.15&1.99-2.30\\
\hline
\multirow{7}*{KE}&J&1.248&1.166-1.330\\
&Hcont&1.5804&1.5688-1.5920\\
&FeII\textsuperscript{1}&1.6455&1.6327-1.6583\\
&CH4s\textsuperscript{1}&1.5923&1.5295-1.6552\\
&H&1.633&1.485-1.781\\
&CH4l\textsuperscript{1}&1.6809&1.6125-1.7493\\
&Kp&2.124&1.948-2.299\\
&Ks&2.146&1.991-2.302\\
&H2\,{\footnotesize v=1-0}\textsuperscript{1,2}&2.1281&2.1112-2.1452\\
&Br\_gamma\textsuperscript{1}&2.1686&2.1523-2.1849\\
&PAH&3.2904&3.2627-3.3182\\
&Lp&3.776&3.426-4.126\\
\hline
\multicolumn{4}{l}{\textsuperscript{1} \small associated with a blocker PK-50.}\\
\multicolumn{4}{l}{\textsuperscript{2} \small called NB2.108  in the fits headers.}\\
\end{tabular}
\caption{\sl  Spectral filter specifications. GN, GS, and KE stand for Gemini North (NIRI), Gemini South (NICI), and Keck II (NIRC2).}   \label{tab: filt}
\end{center}
\end{table}
Images with $\sim30$s exposure time were recorded in an ADI mode. With NIRI, ADI consists of recording a sequence of many exposures of the target while keeping the instrument rotator off. The camera and the telescope optics thus remain aligned for all exposures, enabling a very accurate point spread function (PSF) correlation between consecutive exposures. Since the field of view rotates when the rotator is off, any off-axis source (companion or background object) moves angularly around the central star with time. For each frame of the sequence, a locally optimized combination of images \citep[LOCI,][]{lafreniere07b,marois10} produces a reference PSF to subtract. The PSF subtracted frames are then rotated to align north up and they are median combined. Generally, the more the field of view rotates during the sequence, the more efficient the speckle subtraction. Optimal performance is achieved if the observations are acquired as the target is close to the meridian ($\pm$1h from its transit).

During the first observations of each star (apart from 9 that saturated the detector even for the shortest exposure time), unsaturated short exposures were recorded to calibrate the photometry in the~$\sim$30s images. Sometimes, when observing at a second or later epoch, to save time we did not acquire unsaturated images, observe more targets, and follow more exoplanet candidates.
  
\subsection{NIRC2 at Keck II}
\label{subsec: keck}
Most of the IDPS data were obtained with the NIRC2 instrument~\citep{mclean03,wizinowich06} at Keck II during~$\sim$70 nights or half-nights between~2001 and~2014.

Two observing modes were employed, and both of these modes enabled the use of ADI to subtract the speckle pattern. In the position angle mode~(PA in Tab.\,\ref{tab: obslist0}), north points to the same direction in all frames of a given sequence.  The field of view is thus fixed while most of the speckle pattern rotates. One can thus discriminate a fixed off-axis source from the rotating speckle pattern. In the vertical angle mode~(V), the pupil rotator is kept off and the field rotates, which is equivalent to the ADI mode used at Gemini. The~V mode is usually more effective for speckle calibration, but this mode was not used before~2004 since we were still searching for the best observing strategy at this time. 

We sometimes recorded saturated and unsaturated data but usually, we only recorded long exposure sequences using one of the occulting coronagraphic focal plane masks, with diameters of 400, 600, 800, or 1000\,mas. The masks are not fully opaque and unsaturated images of the star could be extracted from the long exposures. During the survey, we used several filters that are listed in~Tab.\,\ref{tab: filt}.

Most of the data were recorded with the $10''\times10''$ narrow field-of-view camera with 9.942\,mas/pixels. We also used the $40''\times40''$ wide camera with 39.686\,mas/pixels (w in Tab.\,\ref{tab: obslist0}) in peculiar cases to probe planets at more than $10''$ from their star or when the observing conditions where bad.

\subsection{NICI at Gemini South}
Thirty-four stars were observed with the NICI AO camera~\citep{chun08} in~2009, 2012, and 2013~(programs GS-2009B-Q-14, GS-2012B-Q-8, and GS-2013A-Q-24). The camera provided a  $18.5"\times18.5"$ field of view and a 18\,mas plate scale.  We recorded 1024$\times$1024 pixel images in one or two filters at the same time~(Tab.\,\ref{tab: filt}). All data were taken in~ADI mode keeping the Cassegrain rotator off~(V mode). Sometimes, the SDI mode was also used observing simultaneously at CH4s and CH4l. We always used the F0.32~(320mas diameter) or F0.65~(650mas) occulting focal plane masks. As for NIRC2, the occulting masks were not fully opaque and unsaturated images of the star could be extracted from the long exposures.

\subsection{NICMOS on HST and other programs}
We reduced public data that were not part of the IDPS to obtain a second epoch for some of our candidates. We used NIRI, NICI, NIRC2, NACO, and NICMOS archive data. We list the programs these data come from in Tab.\,\ref{tab: obslist0}.

\section{Image processing}
\label{sec:pipeline}
The pipeline we developed for a uniform processing of all the IDPS sequences can be divided into four main steps. First, flat fielding, sky subtraction, bad-pixel and distortion corrections were applied to each frame of a considered sequence (\S\,\ref{subsec: firststep}). In the second step, all frames were registered within a pixel~(\S\,\ref{subsec: center}). The procedures used for these two first steps depended on the instrument and on the sequence (procedures were not the same for registering saturated or unsaturated data for example). The last two steps consisted of obtaining calibrated high contrast images and were common to all the IDPS sequences. In the third step, a LOCI-like algorithm was applied to each sequence to reduce the speckle intensity and provide a final image with high contrast for each target~(\S\,\ref{subsec: highcontr}). In cases where the field-of-view rotation was small, north was aligned up in each frame of the sequence and all the frames were stacked, applying no speckle subtracting algorithm. The fourth step calibrated the contrast level in the final images, estimated the astrometry and the photometry of off-axis sources, and gathered the results in a database~(\S\,\ref{subsec: photo}).

\subsection{Cosmetic corrections}
\label{subsec: firststep}
Considering one IDPS sequence, the first step of our data processing was the selection of the frames. If their quality was degraded because of clouds, high seeing variations, or bad adaptive optics corrections, they were removed from the sequence. When available, a sky frame was subtracted from all frames. In the cases of observations at Lp or Mp bands, the LOCI background subtraction algorithm presented in \citet{galicher11c} was used to subtract the background residuals.

\begin{figure}[!ht]
\centering
    \includegraphics{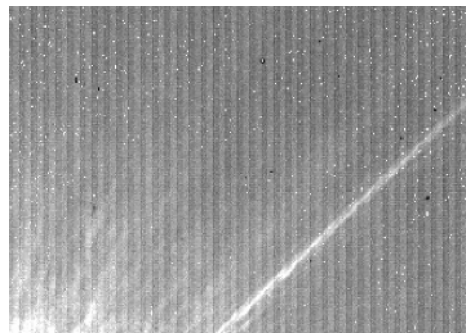}
    \caption{\sl Stripe pattern induced by detectors in NIRI and NICI images.}
    \label{fig: stripe}
  \end{figure}

Each frame of the sky-corrected sequence was corrected from flat-fielding using calibration data taken at the telescope. Then, the bad and hot pixels were found from the sky and flat images, and their flux was set to the median value of neighboring pixels of the science image. Sometimes, detectors created a horizontal or vertical stripe pattern in NIRI and NICI images (Fig.\,\ref{fig: stripe}).
To remove this pattern, we assumed it to be translation invariant.  We estimated the stripe intensity and spatial frequency from the 25 first rows for vertical stripes or columns for horizontal stripes and we subtracted the pattern to the rest of the image.

\begin{figure}[!ht]
\centering
\includegraphics{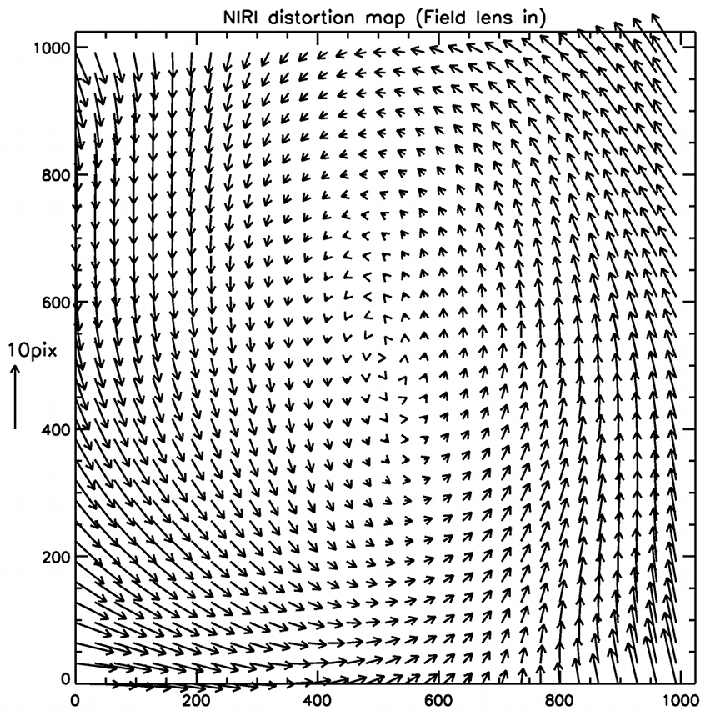}\\
\includegraphics{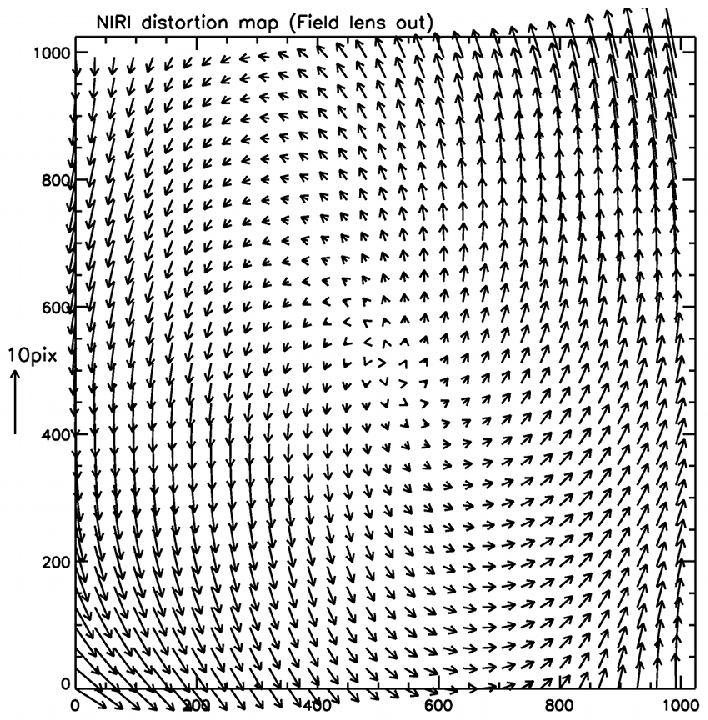}
\caption[]{\sl NIRI distorsion maps for images taken with~(top) or without~(bottom) the Altair field lens. Arrows indicate the distortions at each position of the detector. For the sake of visibility, we multiplied their length by 10.}
\label{fig: plateGN}
\end{figure}

Then, the field distortions were corrected. For the NIRC2 narrow camera images, we used the solution from \citet{yelda10} with an accuracy down to~$\sim$0.01\,pixel (0.5mas) for static distortion correction. When observing with the wide camera, we applied the Keck routine revised by Cameron\,\footnote{\tiny\href{http://www.astro.caltech.edu/~pbc/AO/distortion.pdf}{http://www.astro.caltech.edu/\string~pbc/AO/distortion.pdf}} that corrects for distortions down to a~$\sim$4\,mas level. For~NICI distortions, we used the instrument team solutions that have an accuracy of~$\sim$35\,mas for 2009 data. To correct for NIRI distortions, we compared~NIRI and~NIRC2 images of the Trapezium region and we created new distortion maps for images taken with or without the Altair field lens~(Fig.\,\ref{fig: plateGN}).
In both cases, the accuracy of the correction is less than a pixel~($<$21\,mas) close to the center of the detector but reaches~2 to~5\,pixels~(40-100\,mas) further than~$\sim$4" from the center.  We did not correct for distortions in NICMOS images and assumed a 1.5\,pixel~(113\,mas) error.

Eventually, we registered all images within $\sim1$\,pixels adding zero pixels at the borders of the images to ensure no field-of-view loss when rotating the frames as described in \S\,\ref{subsec: highcontr}.

\subsection{Image registration}
\label{subsec: center}
If images were unsaturated, we used an iterative Gaussian fit procedure to register each frame of the sequence within a fraction of pixel~($\lesssim$1/50). We used the same procedure to register coronagraphic images since the occulting mask transmissions was partially transmissive and left unsaturated images of the central star on the detector. In NIRC2 and NICI coronagraphic images, the central part of the star image, which was attenuated by the focal plane mask, was slightly shifted with respect to the external part of the star image (i.e., the speckles that were not affected by the focal plane mask). We thus registered all the frames using the deflected central star image and compensated for the bias induced by the mask. Several times during the IDPS, we calibrated the mask deflections at Keck II observing the HD\,172649 binary system with the~400, 600, 800, and 1000\,mas diameter masks at~J, H, and Ks filters.

For NIRI saturated images, we selected the first frame of the sequence and we masked all the pixels but the regions of the spider diffraction pattern. We then found the position that maximized the correlation of the masked image with its~180 degree rotation~(spiders are symmetrical). Once the first frame was registered within a pixel, we built a new mask to select the spider diffraction pattern and an annulus where speckles were not saturated and where the detector noise was negligible. We registered all the frames of the sequence maximizing their correlation inside the new mask with the first frame. Sizes of the annulus were sequence dependent.

For NIRC2 saturated images, the procedure was very similar but the centering of the first frame was done by hand~($\sim$1pix error) using the hexagonal diffraction signal as a guide because the spider pattern was not detected.

For NICMOS images, we selected the spider pattern area and found the position that maximized the correlation of the image with its~180 degree rotation.

At this point of the pipeline, all sequences were registered within $\le$1\,pixel.

\subsection{High contrast imaging}
\label{subsec: highcontr}
\begin{figure}[!ht]
\centering
\includegraphics{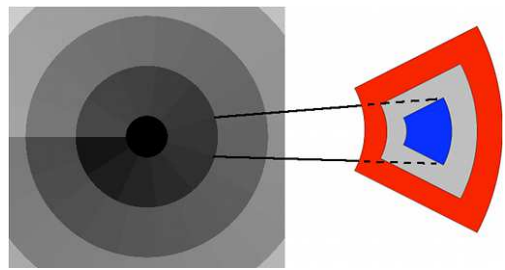}
\caption[]{\sl SOSIE sections of interest~(left and right central blue section) and optimizing region~(right exterior red region).}
\label{fig: locisec}
\end{figure}

For NIRC2 and NIRI data, the speckle pattern was reduced in each frame of the sequence by subtracting an optimized reference image calculated by the SOSIE algorithm~\citep{marois10}. To create the reference image, we divided the target image into sections~(left in Fig.~\ref{fig: locisec}). For each section~(blue central region, right in Fig.~\ref{fig: locisec}), we selected the frames in the sequence for which the field-of-view rotation would have sufficiently displaced a putative companion~(half of the~PSF full-width half-maximum, FWHM), and we defined a region~(exterior red region) setting a FWHM gap. We found the linear combination of the selected frames that minimized the residual noise in the exterior region.
We then subtracted the linear combination in the section of interest~(blue central section). This reduced the speckle intensity because the SOSIE coefficients that are needed to subtract the speckles in the blue region are highly correlated with the SOSIE coefficients that are optimized to subtract the speckles in the red region. The FWHM gap avoids self-subtraction of the planet signal. We tested several geometrical shapes for the sections and found that annuli were a good trade-off between computing time and performance~(negligible loss in performance). After subtraction, we derived the north axis from the~FITS header keywords for each frame and we rotated all frames to align their field of view with north up. We eventually median combined the frames to enhance the signal-to-noise ratio of putative off-axis sources. In the regions of the image where no reference frames were available to apply the SOSIE algorithm~(far from the central star if the telescope was dithered during the sequence or close to the center when the field-of-view rotation was small), we did no speckle subtraction and only median combined the frames.

For sequences taken in PA mode~(\S\,\ref{subsec: keck}), the pupil rotated between frames rather than the field of view. We rotated each frame to create a sequence with a fixed speckle pattern and a rotating field of view. We then applied our SOSIE algorithm.

When only one spectral filter was used with~NICI, we used the same SOSIE algorithm as described before. For images taken at two spectral bands simultaneously, we developed a SOSIE-based algorithm that minimizes the speckle intensity using the spectral images~(spectral speckle differential imaging) and the~ADI images simultaneously \citep[see also][]{artigau08}.  This data processing was used only for detection.  To avoid the issue of calibration in such ADI/SDI SOSIE images~\citep{maireAL14}, we used the ADI SOSIE images (no SDI) to derive the photometry of the candidates and the detection limits.

Usually, NICMOS sequences are composed of only 2-6 frames, which limits the SOSIE performance. To improve the speckle subtraction, we built a NICMOS image database (called $I_\mathrm{DB}$) gathering~$1012$ images that is similar to what \citet{lafreniere09} proposed. To reduce the speckle level in one section of one frame, we selected the frames of the same sequence and the 50 most correlated images $I_\mathrm{DB}$ in this section. Then, we applied the SOSIE algorithm described previously.

As the~ADI observing mode was not used before 2004, the SOSIE algorithm could not be applied to a non-negligible number of sequences taken between 2000 and 2004 with NIRC2. For these sequences, we only rotated all the frames to align the north up and we median combined them. We tested a SOSIE algorithm using a library of PSFs taken the same night with the same instrumental configuration (similar to the NICMOS-HST data processing) but no improvement was achieved.  The reason is certainly that aberrations vary over the night in a ground-based telescope (pointing, flexures, temperatures, etc.) and thus, the PSF and the speckle pattern evolve over the night.

\subsection{Photometry and astrometry calibrations}
\label{subsec: photo}
\begin{table*}[!ht]
\begin{center}
\begin{tabular}{ccccccccc|ccc}
\hline
Instrument&\multicolumn{8}{c}{NIRC2}&\multicolumn{3}{|c}{NICI}\\
Filter&J&H&H&H&Ks&Ks&Ks&Lp&CH4l&H&K\\
Mask diameter (mas)&800&600&800&1000&400&600&800&800&320&320&320\\
Attenuation&70&200&40&39&57&29&28&120&300&238&191\\
\end{tabular}
\caption{\sl Intensity attenuation of the PSF behind the NIRC2 and NICI occulting masks.}
\label{tab: filtcoro}
\end{center} 
\end{table*}

\begin{table*}[!ht]
\begin{center}
\tiny
\begin{tabular}{lccccccccccccccc}
 Star&Date&Filter&0.3"&0.5"&1.0"&1.5"&2.0"&3.0"&4.0"&5.0"&6.0"&7.0"&8.0"&9.0"&10.0"\\
\hline
HR9&2009-11-01&Ks&5.3&6.9&9.7&11.6&12.8&13.5&13.6&13.4&12.6&9.8&-&-&-\\
HR9&2009-12-05&CH4l&8.5&10.0&12.7&13.9&14.7&15.2&15.3&15.3&15.3&15.3&15.7&16.6&15.1\\
\end{tabular}
\caption{\sl $5\,\sigma$ contrast curves of the IDPS observations. The complete list is given in Tab.\,\ref{tab: contrlist}.}
\label{tab: contrlist0}
\end{center}
\end{table*}

As for any~ADI or SDI technique used for speckle calibration, the astrometry and photometry are biased in the final images. The procedure we used to estimate these biases requires an unsaturated~PSF~\citep{galicher11b}. When unsaturated images were available, they were registered following the procedures described in \S\,\ref{subsec: center} and we median combined them to obtain the needed unsaturated~PSF.

In the case of NIRC2 coronagraphic images, we used the attenuated images of the star that were recorded behind the occulting mask. We compensated for the mask transmissions that were calibrated observing binary systems~(\S\,\ref{subsec: center} and Tab.\,\ref{tab: filtcoro}). For the 600 and 800\,mas masks at H and Lp, we used images of GJ\,803 and HR\,8799 taken with and without masks for the calibration.

All the NICI data were recorded with an occulting mask and we also used the attenuated image of the star behind the coronagraph as a reference for photometry.  From the NICI documentation, the attenuation of the 0.32 arcsec occulting mask was~$\sim$300. We used this value for the CH4 bands as the calibration was performed at CH4l.  For the K and H bands, we used calibrations published in~\citet{wahhaj11} in which the attenuations were 238 and 191 at H and K, respectively.

For several data sets, saturated and unsaturated sequences were not recorded using the same filters although their central wavelengths and bandwidths were close. We thus considered the median saturated and unsaturated images. We selected an annulus in which the saturated image was not saturated and where the unsaturated image was not dominated by the detector noise. We integrated the flux over the annulus for both images and we determined the factor that gives the photometry in the saturated image from the measured flux in the unsaturated image. We finally obtained an unsaturated~PSF that was calibrated in flux with respect to the saturated data.  Doing so, we assumed the PSF pattern was similar in the two filters, which was a good approximation as the filter central wavelengths and bandwidths were similar.

\begin{figure}[!ht]
\centering
\includegraphics{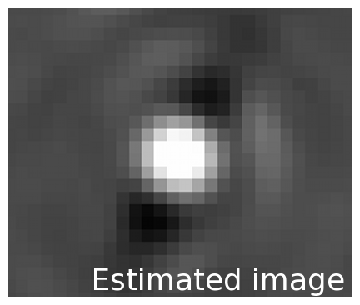}\hspace{0.2cm}\includegraphics{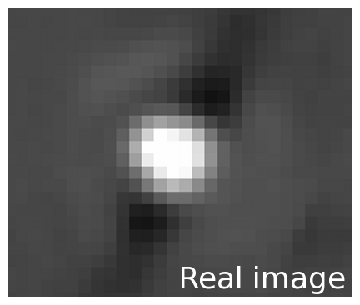}
\caption[]{\sl Estimated~(left) and real~(right) images of an off-axis source.}
\label{fig: template}
\end{figure}

When a binary or planet candidate was detected, we used the calibrated unsaturated~PSF to estimate its astrometry and photometry~\citep{galicher11b}. First, we roughly estimated the flux and position~(within~1 pixel) of the source in the~SOSIE image. The pipeline then created a data cube of frames that only contained the unsaturated~PSF at the candidate position on the detector accounting for the field-of-view rotation in each frame and for the smearing of off-axis source images as the field of view could rotate during a single exposure. The SOSIE coefficients that were used to obtain the~SOSIE image where the candidate was detected were applied on the candidate data cube. The resulting frames were rotated to align north up. The median of these frames provided the estimation of the candidate image in the SOSIE image~(Fig.\,\ref{fig: template}).
We then adjusted the estimated image subpixel position and its flux to minimize the integrated flux of the difference between the real and estimated candidate images. We used a~3\,FWHM diameter disk for the minimization. The~1\,$\sigma$ error bars were the required excursions in position or in flux to increase the minimum residual flux by a factor of~$\sqrt{2}$. We empirically determined this factor running tests on sequences in which we injected known fake planets.

Using the calibrated unsaturated~PSF, we also estimated the SOSIE throughput in all SOSIE sections~(central blue section in Fig.\,\ref{fig: locisec}) following a procedure similar to the one used for the candidate position and flux estimation. The image were thus flux calibrated.

\begin{figure*}[!ht]
\centering
\includegraphics[width=.35\textwidth]{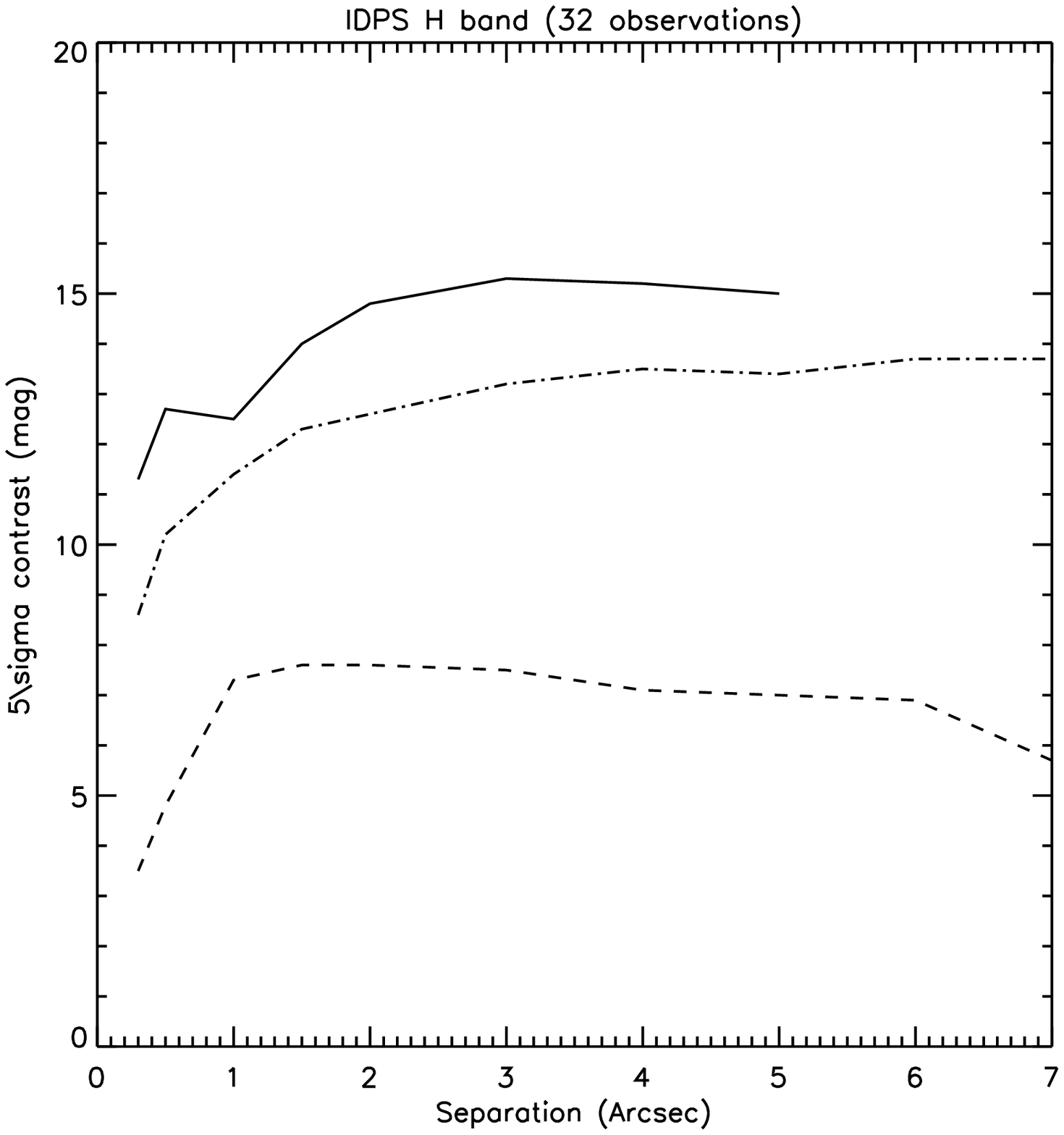}\includegraphics[width=.35\textwidth]{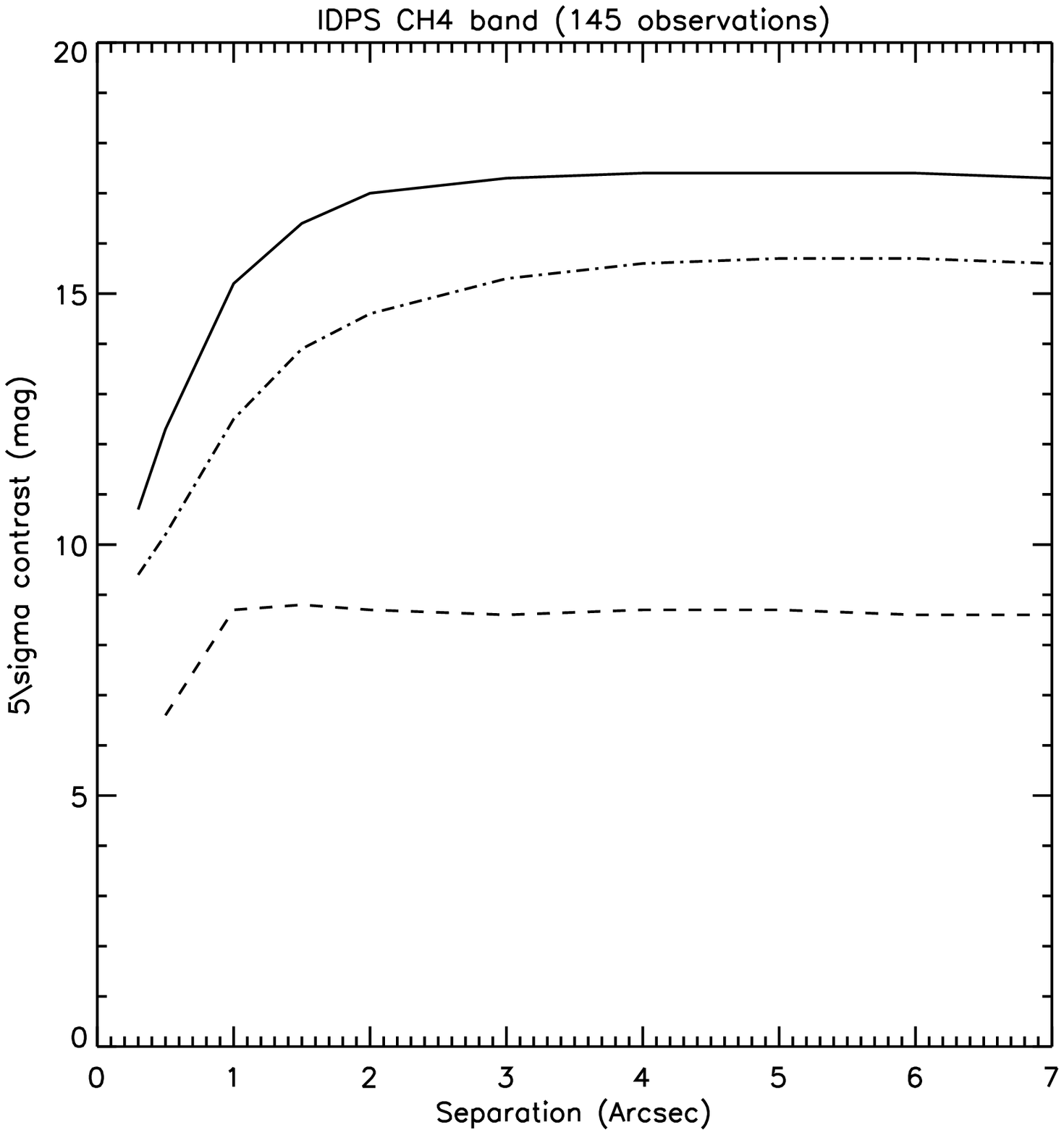}\\\includegraphics[width=.35\textwidth]{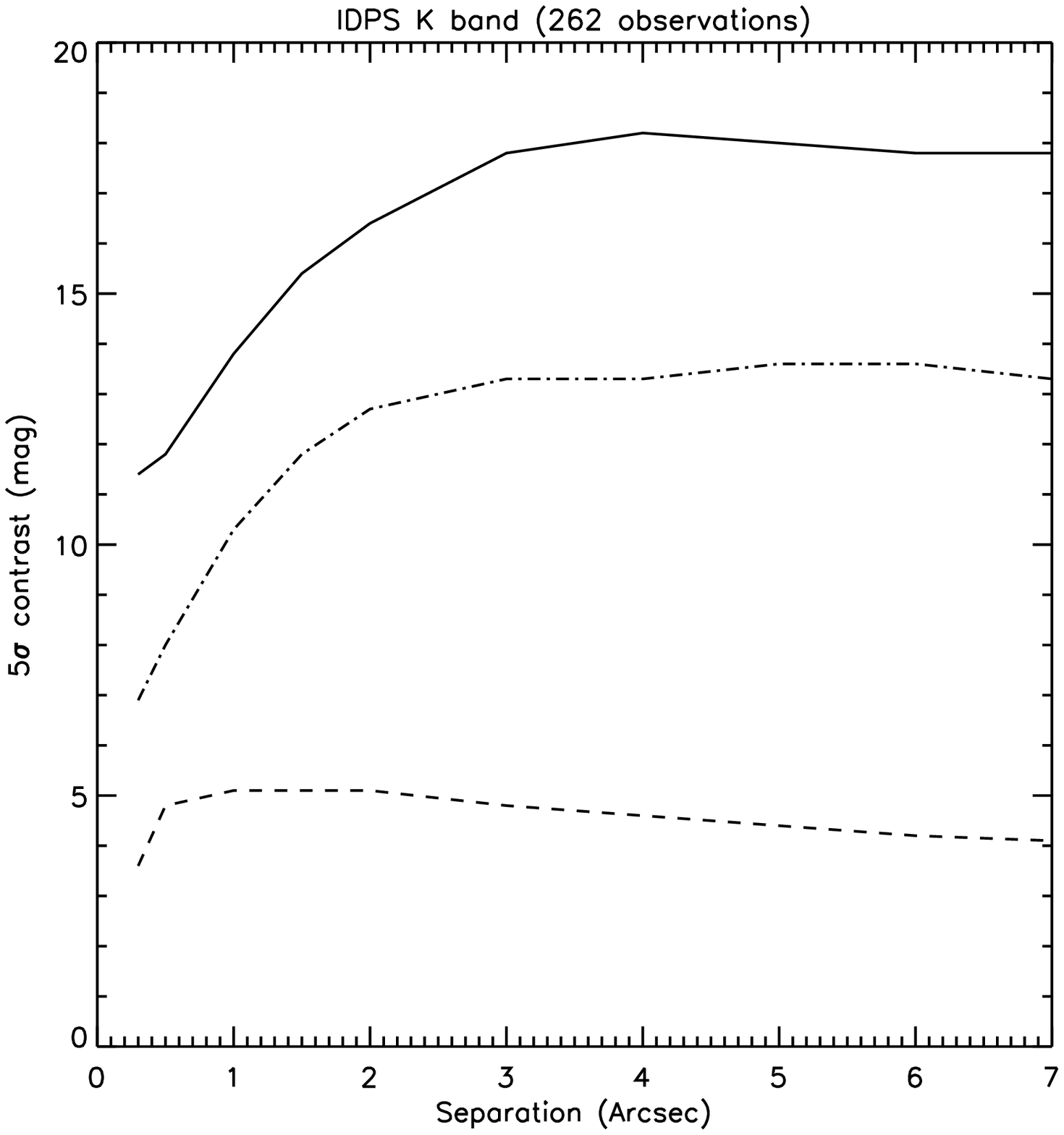}\includegraphics[width=.35\textwidth]{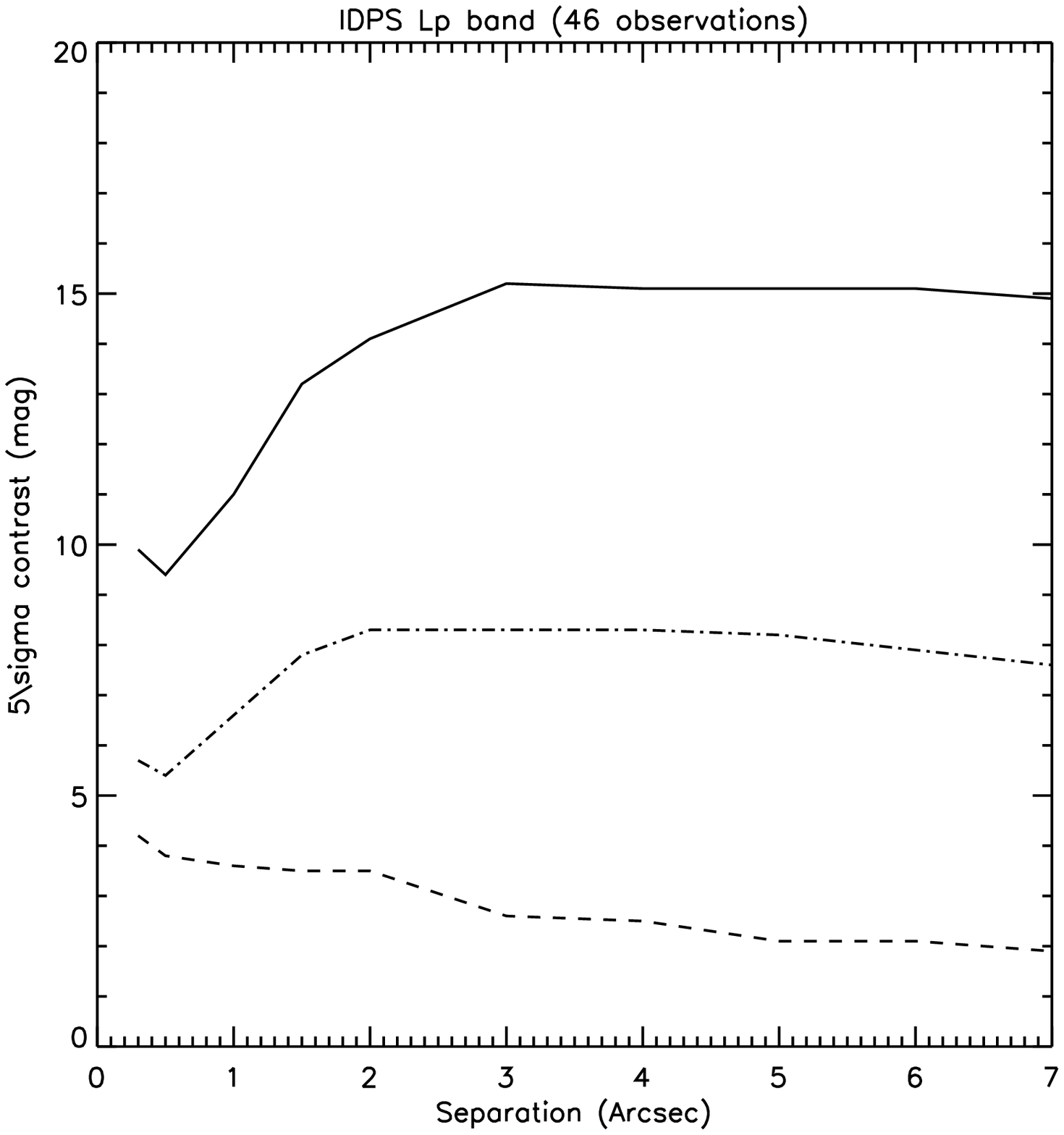}
\caption[]{\sl Typical deep~(full line), median~(dotted-dashed line), and shallow~(dashed line) IDPS $5\,\sigma$ contrast curves against angular separation. Top left: H band. Top right: CH4 bands. Bottom left: K band. Bottom right: L band.}
\label{fig: contrast}
\end{figure*}
For every star (IDPS and public data) with unsaturated data, we produced a contrast curve. When off-axis sources were detected in the image, we added a negative object at their position in the raw datacube to remove them from the image and not bias the contrast at their angular separation~\citep{marois10b,galicher11b}. The noise distribution in ADI processed images is well reproduced by a Gaussian function \citep{marois08b} and we estimated the $5\,\sigma$ noise level where $\sigma$ is the azymuthal standard deviation of the residual flux in annulii of $1\,\lambda/D$ width rejecting pixels with no flux (out of the field of view). We did not account for the small amount of noise realizations \citep{mawet14} whose impact is negligible at the considered separations. Finally, the $5\,\sigma$ noise levels were divided by the stellar flux estimated from the unsaturated images. Typical contrast curves of the IDPS are plotted in Fig.\,\ref{fig: contrast}. The shallow curves correspond to the short sequences taken between~2000 and 2004 with NIRC2. The best performance is commonly reached using NIRC2~(after 2004) or NICI. All the IDPS contrast curves are gathered in~Tab.\,\ref{tab: contrlist0}.

\section{Detections}
\label{sec:results}
\subsection{Crowded fields}
\label{subsec: crowded}
For~$5$ targets, we detected more than 50 off-axis sources~(Tab.\,\ref{tab : crowded}). These stars are within 10\,deg in latitude from the Galactic plane and the probability that the off-axis sources are background objects is high. Thus, we put a low priority on these stars, and finally, we did not reobserved them.
\begin{table*}[!ht]
\begin{center}
\begin{tabular}{ccccc}
\hline
Name&$\alpha$&$\delta$&Date\\
&(J2000.0)&(J2000.0)\\
\hline
2MASSJ17150362-2749397&17:15:04&-27:49:40&2010-07-12\\
HD324741&17:54:55&-26:49:42&2004-05-24\\
HIP89829&18:19:52&-29:16:33&2009-08-31\\
HIP93805&19:06:15&-04:52:57&2008-05-31\\
HIP99770&20:14:32&+36:48:23&2008-05-22\\
\hline
\end{tabular}
\caption{\sl Targets with crowded field (more than $50$ sources).}
\label{tab : crowded}
\end{center}
\end{table*}

\subsection{Binary and triple systems}
\begin{figure}[!ht]
\centering
\includegraphics[width=.43\textwidth]{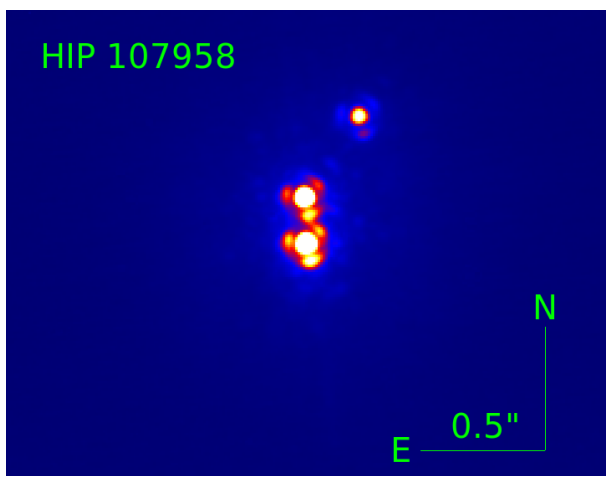}
\caption[]{\sl Multistellar system discovered during the IDPS: HIP\,$107948$.}
\label{fig: hip107}
\end{figure}

During the survey, we detected $59$ visual multiple systems with projected separations smaller than 200\,AU. Forty of these systems are already known physical binaries and $16$ are physical binaries discovered during the IDPS. We also obtained an image of HIP\,$117452$, which is a triple stellar system~\citep{derosa11}. Finally, we discovered two triple stars: HIP\,$104365$ \citep{vigan12} and HIP\,$107948$ (Fig.\,\ref{fig: hip107}).
The properties of the binaries (separation, magnitude difference, etc.) are listed in~Tab.\,\ref{tab: bin0}. We do not derive the stellar type of the companions in this paper.
\begin{table*}[!ht]
\begin{center}
\tiny
\begin{tabular}{lllcccccccc}
Star&$\alpha$&$\delta$&$\rho$&$\theta$&Contrast&Filter&Date&Telescope&Status&Comments/References\\
&(J2000.0)&(J2000.0)&(arcsec)&(deg)&($\Delta$mag)&&&&&\\
\hline
TYC1186-706-1&00:23:34.66&+20:14:28.75&1.74$\pm$0.01&136.7&0.6&Kcont&2008-12-17&KE&A\textsuperscript{a}&\\
&&&1.74$\pm$0.01&137.4&0.6&Kp&2009-07-30&KE&A\textsuperscript{a}&\\
\hline
\multicolumn{11}{l}{\sl \textsuperscript{a,b,c} Companion status confirmed from the IDPS, from previous works, or combining the IDPS and previous works (reference column) respectively.}\\
\end{tabular}
\caption{\sl Binaries detected during the IDPS. In the status column, 'A' stands for associated with the star and 'B' for background object. '?' means unknown status. The complete table is given in Tab.\,\ref{tab: bin}.}
\label{tab: bin0}
\end{center}
\end{table*}

\subsection{Exoplanet detections}
\begin{figure}[!ht]
\centering
\includegraphics[width=.43\textwidth]{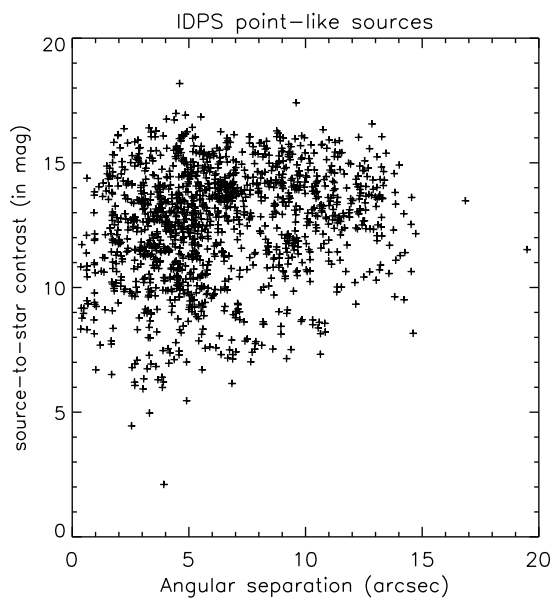}
\caption[]{\sl Contrast of 1,396 point-like sources detected in the IDPS as a function of their angular separation.}
\label{fig: contrsepaIDPSsource}
\end{figure}

\begin{figure}[!ht]
\centering
\includegraphics{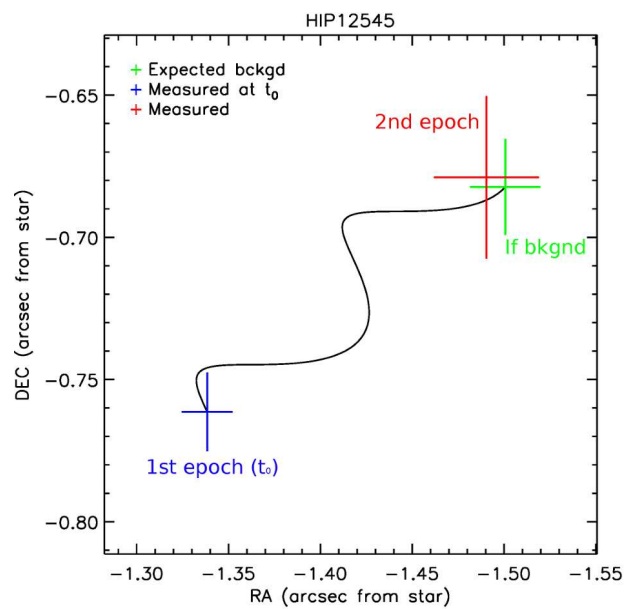}
\caption[]{\sl Example of the astrometry test for comparing the expected position of a background source (blue and green) to the measured positions (blue and red) for a planet candidate around HIP\,12545. Crosses include error bars on the measured astrometry and on the stellar parallax and proper motion.}
\label{fig: astrotest}
\end{figure}

One of the main results of the IDPS is the discovery of the four exoplanets orbiting the HR\,8799 star \citep{marois08,marois10}. We also flagged $2,279$ point-like objects in the IDPS images, corresponding to $\sim1,100$ individual sources detected at 2-3 epochs. We plot in Fig.\,\ref{fig: contrsepaIDPSsource} the contrast and the angular separation with respect to the star for the 1,396 IDPS detections whose contrast could be estimated.
By observing a given star at multiple epochs and considering proper motions and parallax, we confirmed that most of the detected sources are background objects (Fig.\,\ref{fig: astrotest}).
Some candidates have not yet been followed up because of very bad weather or technical problems. The characteristics of the $2,279$ detected sources are reported in Tab.\,\ref{tab: cand0} and are available on a dedicated website\footnote{\tiny\href{http://lesia.obspm.fr/directimaging/admin/form_user.php}{http://lesia.obspm.fr/directimaging/admin/form\_user.php}}.
\begin{table*}[!ht]
\begin{center}
\tiny
 \resizebox{\hsize}{!}{
\begin{tabular}{lcccccccccccccc}
Star&$\alpha$&$\delta$&$\rho$&$\theta$&$\delta\rho=\rho\delta\theta$&$\Delta m$&Filter&$M$&$m_{\mathrm{min,max}}$&$a_\mathrm{proj}$&Date&Tel.&\#&Status\\
&(J2000.0)&(J2000.0)&(")&(deg)&(mas)&&&&($M_\mathrm{Jup}$)&(AU)&&&\\
\hline
HR9&00:06:50.09&-23:06:27.14&19.691&337.347&40&--&Kp&--&--&770.1$\pm$13.0&2001-12-01&KE&1&?\\
HIP682&00:08:25.74&+06:37:00.48&5.472&268.037&14&15.47$\pm$0.49&Ks&--&--&--&2012-09-04&KE&1&B\\
&&&4.945&267.774&12&14.73$\pm$0.31&Kp&--&--&--&2006-07-18&KE&1&B\\
\end{tabular}}
\caption{\sl substellar candidates and background sources observed during the IDPS.  In the status column, 'A' stands for associated with the star, whereas 'B' is for background object and 'F' for false detection.  The flag 'B?' means that the errors bars are larger than parallax and proper motion, but given the displacement of all the candidates between the two epochs, we are confident the object is background.}
\label{tab: cand0}
\end{center}
\end{table*}

\section{Occurence of stars with giant planets}
\label{sec: plafreq}
The detection limits (Tab.\,\ref{tab: contrlist0}) and the four exoplanet detections of the IDPS can be used to run a Monte Carlo statistical analysis (\S\,\ref{subsec: formMC} and \S\,\ref{subsec: plaprob}) and calculate an upper limit to the frequency of stars that harbor at least one companion of mass and semimajor axis within given intervals (\S\,\ref{subsec: uplimit}). Moreover, as exoplanets were discovered during the survey, we can put a lower limit on this frequency and we study the impact of the hosting star mass (\S\,\ref{subsub: uniform}). We also study the impact of one of the assumptions that is needed to run the Monte Carlo analysis (\S\,\ref{subsub: powerlaw}). Finally, we compare our results to previous works (\S\,\ref{subsec: others}).

\subsection{Sample}
\begin{figure}[!ht]
\centering
\includegraphics[width=6.cm]{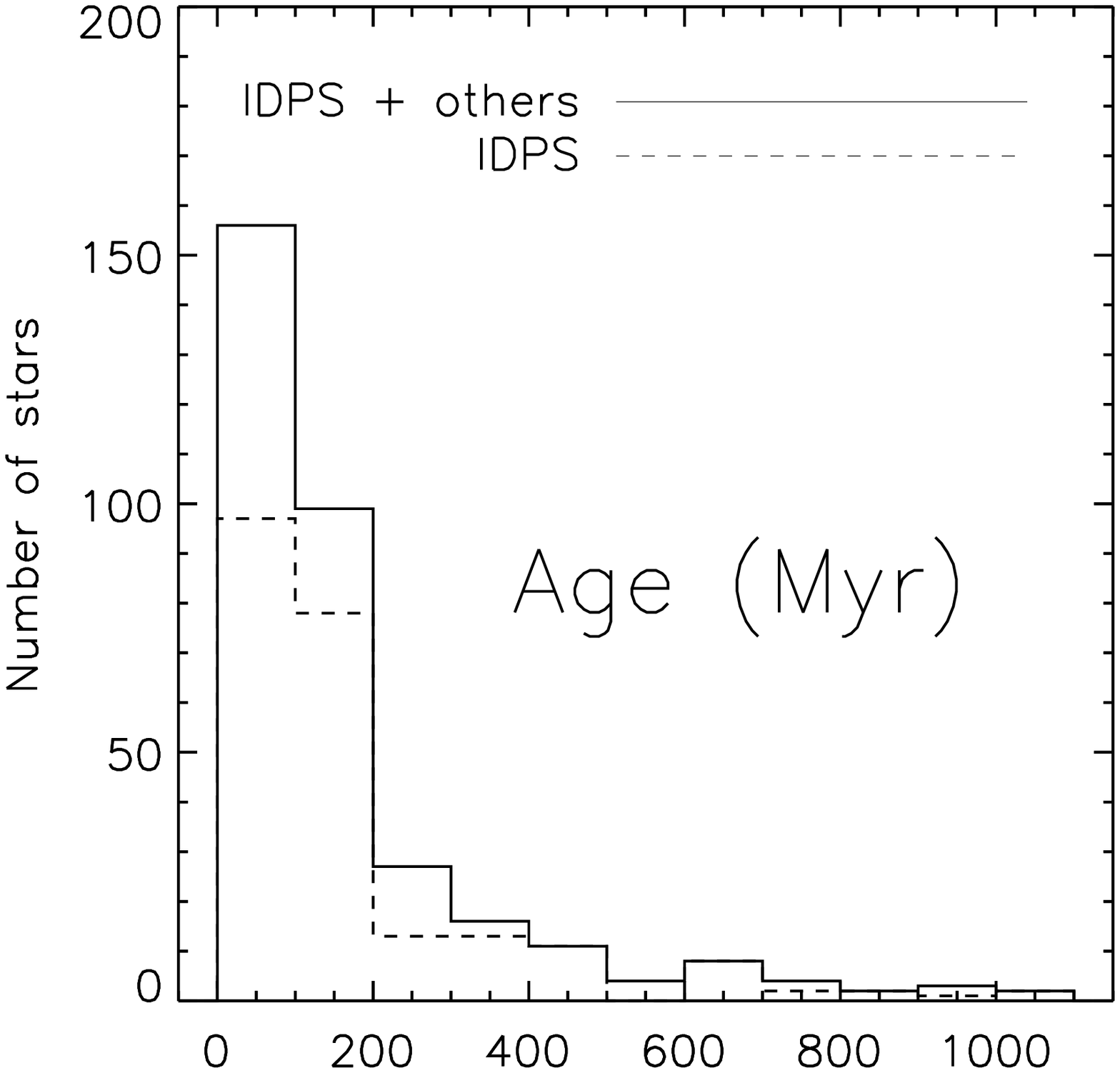}\\
\includegraphics[width=6.cm]{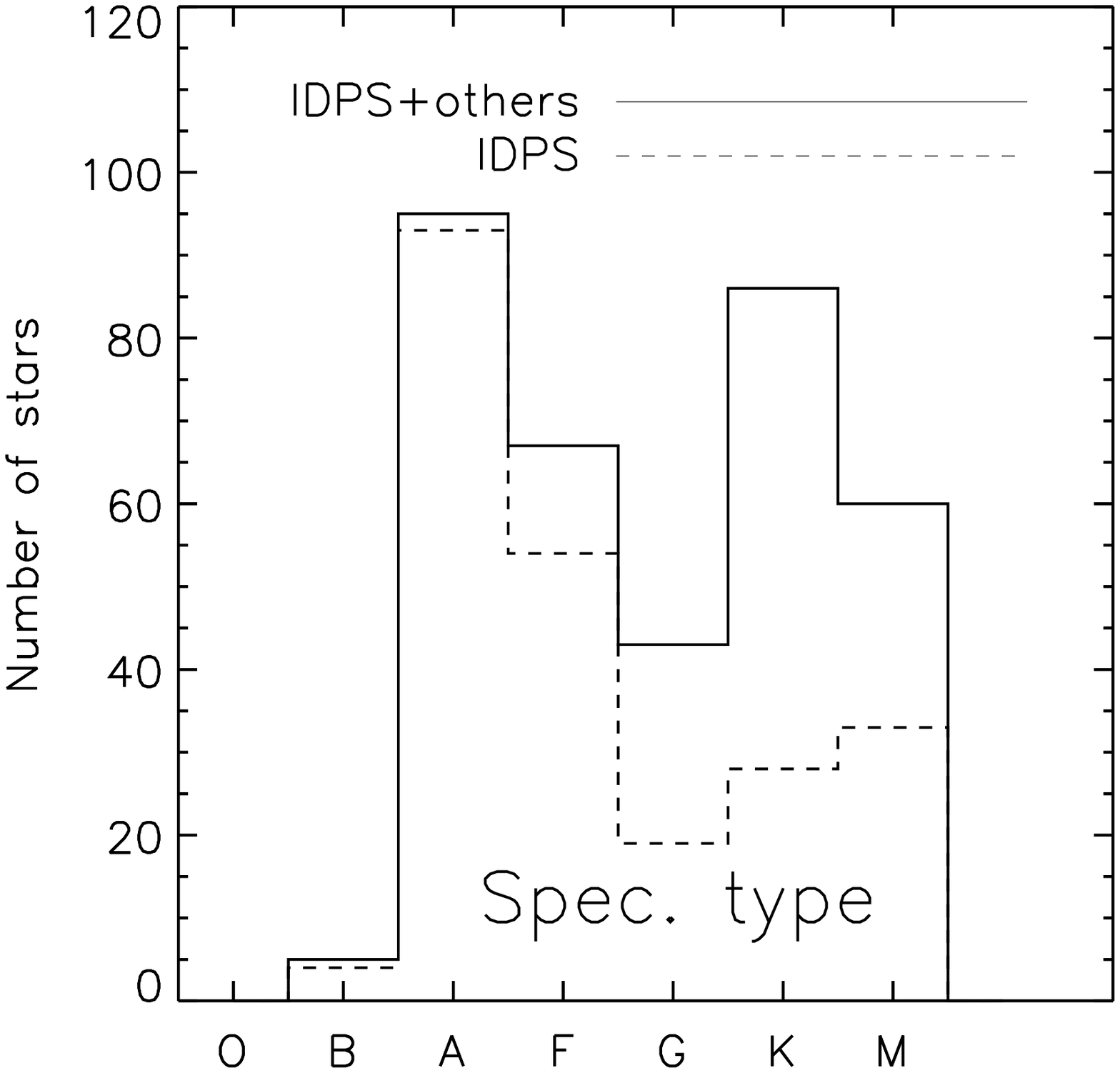}\\
\includegraphics[width=6.cm]{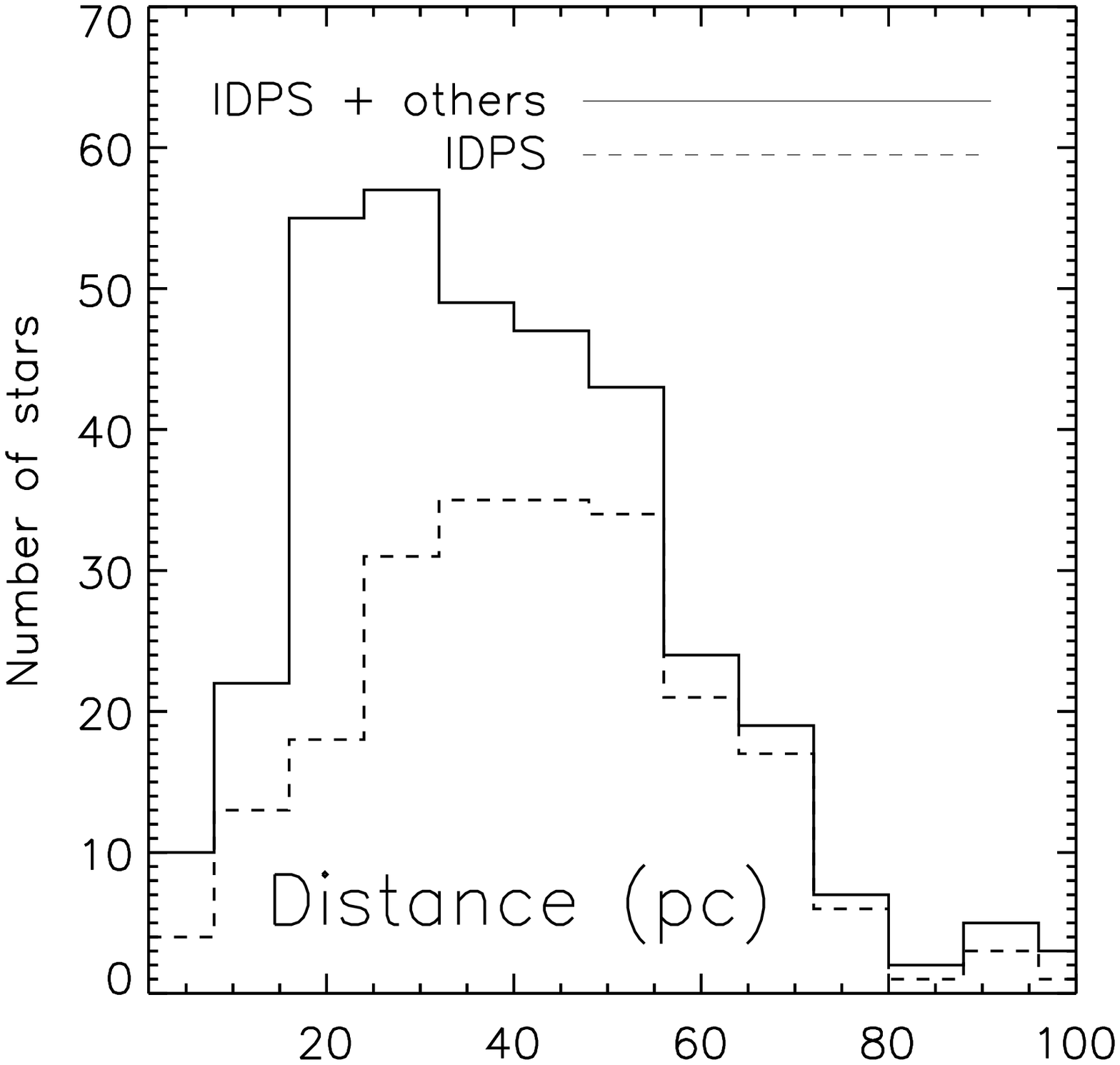}
\caption[]{\sl Distribution of the ages, spectral types, and distances of the IDPS stars with contrast curves (dashed line) and of the sample used for the statistical study (full line).}
\label{fig: starsample_stat}
\end{figure}

We focus our analysis on single star systems rejecting the binaries (Tab.\,\ref{tab: bin0}) from the IDPS sample. We also reject the stars with crowded fields (\S\,\ref{subsec: crowded}) and the stars for which no contrast curve could be derived from the observations. This leaves $229$ stars with a median age of $100$\,Myr and a median distance of $43$\,pc (dashed lines in Fig.\,\ref{fig: starsample_stat}).

 To increase the number of stars in the sample and, thus, derive a more accurate frequency of stars with giant planets, we combine the IDPS detection limits and detections to the detection limits of two other surveys: Gemini deep planet survey \citep{lafreniere07a} and a NACO/VLT survey \citep{chauvin10}. Thanks to previous collaborations with these teams, we know that our detection limits and theirs are consistent. In future works, it would be interesting to combine all the existing surveys ensuring all the detection limits are consistent but this is out of the scope of our paper.

The \citet{lafreniere07a} and \citet{chauvin10} surveys complete the IDPS because they include more G, K, and M stars (full lines in Fig.\,\ref{fig: starsample_stat}).
The complete sample of 356 stars includes 5, 95, 67, 43, 86, and 60 B, A, F, G, K, M stars. The median age and median distance, which are $100$\,Myr and $37$\,pc, are very similar to those of the IDPS. We use this sample in the rest of the paper.

\subsection{Formalism and assumptions}
\label{subsec: formMC}
The analysis is based on the statistical formalism presented in \citet{lafreniere07a}. A Monte Carlo simulation is used to create an ensemble of fake planets computing their projected angular separation with respect to their star assuming given distributions of planetary masses and orbital parameters. Then, considering a model of planet atmosphere and the age of the targets, the fake planet fluxes are derived and compared to the detection limits of the survey to determine what fraction of planets should have been detected. Finally, the frequency of stars that harbor at least one exoplanet is derived using the Bayes' theorem. In this section, we briefly outline the main steps of the analysis specifying the assumptions.

We simulate $N_{\mathrm{pla}}$ fake planets for each target star. First, we draw $N_{\mathrm{pla}}$ semimajor axis values considering a uniform distribution in the interval $[a_{\mathrm{min}}, a_{\mathrm{max}}]$. These values are converted to planet-star separations following the method of \citet{brandeker06} and assuming the orbital eccentricity distribution derived by \citet{kipping13} from the radial velocity exoplanet sample, which is a Beta distribution with parameters a=0.867 and b=3.03. The projected physical separations are converted into angular separations based on the distance of the primary star. Then, we draw $N_{\mathrm{pla}}$ masses following a uniform distribution in the interval $[m_{\mathrm{min}}, m_{\mathrm{max}}]$. Each planet mass is converted into flux in the observing spectral band using the COND03 model \citep{baraffe03} and a random age assuming an asymmetric Gaussian distribution for which the $1\sigma$ lower age, $1\sigma$ upper age, and average age are given in Tab.\,\ref{tab: targlist0}. The planet-to-star contrast is then derived from the stellar magnitude. As we compare this contrast to a 1D-contrast curve, we add a random noise to account for the variations of the noise in the 2D-reduced image. The random noise is drawn from a Gaussian distribution of standard deviation that is equal to the noise measured at the separation of the fake planet. Finally, we derive the probability $p_j$ of detecting a given planet around the $j$th star from the fraction of fake planets lying above the $5\,\sigma$ detection limit.

When the star was observed several times, we combine the detection limits (Tab.\,\ref{tab: contrlist0}) taking the deepest contrast at each separation. Doing so, we assume the putative planets do not move in the image from one observation to the others. During the IDPS, the earliest and latest observations of a given star are separated by $3.5$\,yr in average. Considering the median distance to the stars ($\sim40$\,pc) and the average minimum angular separation of the observations ($0.3$\,''), the closest detectable planets would have a semimajor axis of $12$\,AU and, thus, an orbital period of $\sim30$\,yr for the massive stars ($2$\,M$_\odot$). The putative planets would then have moved by $\sim20$\,pixels in Keck data and $10$\,pixels in Gemini data between the two observations. Such a motion is small enough so that we can combine the detection limits.

\subsection{Exoplanet detection probability}
\label{subsec: plaprob}
\begin{figure}[!ht]
\centering
\includegraphics[width=.43\textwidth]{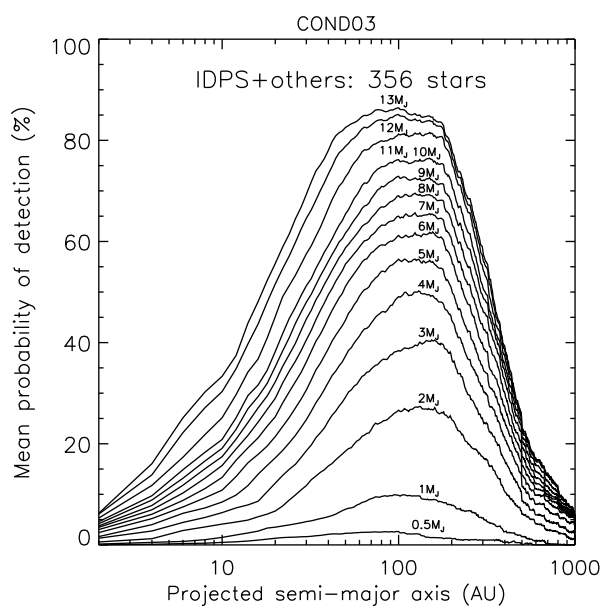}
\caption[]{\sl Mean probability of detecting giant planets during the surveys as a function of the projected semimajor axis for several planetary masses.}
\label{fig: probsma}
\end{figure}

The average of $p_j$ over all $j$ values (i.e., over the complete sample) are plotted in Fig.\,\ref{fig: probsma} as a function of the projected planet-star separation for planet masses going from $0.5$ to $13$ Jupiter masses. For each curve, we set the mass of the $N_{\mathrm{pla}}=100,000$ random planets ($m_{\mathrm{min}}=m_{\mathrm{max}}$).
The peak sensitivity of the survey occurs for projected separations between 20 and 300\,AU. For example, if each star of the sample harbored one $10$\,M$_{\mathrm{J}}$ at 100\,AU, we would have detected $75\%$ of them. The sensitivity declines for separations larger than 200\,AU because of the median distance of the targets ($37$\,pc) and the $\sim10''$ field of view of NIRC2 (the most used instrument during the IDPS).

\subsection{Upper limit on the frequency of stars with giant planets}
\label{subsec: uplimit}
\begin{figure}[!ht]
\centering
\includegraphics[width=.43\textwidth]{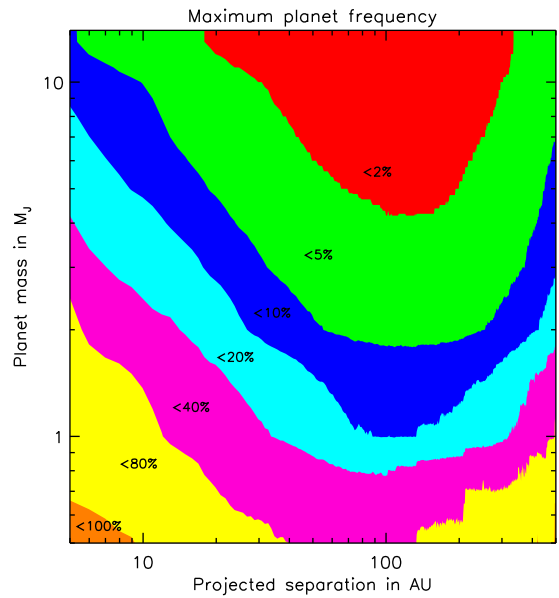}
\caption[]{\sl Upper limit on the frequency of stars harboring at least one giant planet as a function of the planet mass and projected semimajor axis.}
\label{fig: uplimit}
\end{figure}

Using the Bayes theorem, the determination of the $p_j$ yields a credible interval for the frequency of stars with giant planets under the assumptions made about the planet mass and semimajor axis (uniform distributions), stellar age, and atmosphere model. The boundaries of the interval depend on the level of credibility $\alpha$ that is considered \citep[Eq.6 in][]{lafreniere07a}. For $\alpha=95\%$, Fig.\,\ref{fig: uplimit} presents the upper limit on the frequency of stars harboring at least one giant planet as a function of the planet mass and the projected separation.
The large number of targets and the low detection limits of the IDPS, combined with \citet{lafreniere07a} and \citet{chauvin10} results, enables us to constrain the frequency of stars harboring exoplanets with masses down to $0.5-1$\,M$_\mathrm{J}$. For example, we find that there are less than $20\%$ of stars harboring one or more $1$\,M$_\mathrm{J}$ exoplanets between $30$ and $300$\,AU; and less than $10\%$ of stars with one or more $2$\,M$_\mathrm{J}$ exoplanets between $30$ and $400$\,AU.

\subsection{Frequency of stars with giant planets}
\label{subsec: plafreq}
Because substellar objects are detected around stars in our sample~(Tab.\,\ref{tab: detection}), we can also put a lower limit on the frequency of stars with giant planets following the formalism presented in \citet[Eq.6]{lafreniere07a}.
\begin{table}[!ht]
\begin{center}
\begin{tabular}{lc|cccc}
\multicolumn{2}{c}{Star}&\multicolumn{3}{|c}{Planet}\\
name&type&name&mass ($M_J$)&sep. (AU)&Ref.\\
HR\,8799&A5V&\,b&5-11&68&1\\
&&\,c&5-8&38&1\\
&&\,d&5-10&24&1\\
&&\,e&5-10&15&2\\
HD\,130948&G1V&\,b&13-75&47&3\\
HIP\,30034&K2&\,b&13-14&260&4\\
\end{tabular}
\caption{\sl Detected substellar objects in our sample. 1:\citet{marois08}, 2:\citet{marois10}, 3:\citet{lafreniere07a}, 4:\citet{chauvin10}.}
\label{tab: detection}
\end{center}
\end{table}
When calculating the frequency of stars harboring at least one giant planet within $[m_{\mathrm{min}}, m_{\mathrm{max}}]$ and $[a_{\mathrm{min}}, a_{\mathrm{max}}]$, only the exoplanets of Table\,\ref{tab: detection}, whose mass and separation are included in these intervals are considered detected during the survey. For example, the detected exoplanets for $[m_{\mathrm{min}}, m_{\mathrm{max}}]=[0.5,14]\,M_J$ and $[a_{\mathrm{min}}, a_{\mathrm{max}}]=[20,300]$\,AU are HR\,8799\,b, c, d,  and HIP\,30034\,b. Also, even if several of the HR\,8799 planets are flagged as detected, the HR\,8799 system counts for one detection only, i.e., one star with at least one giant planet.

\subsubsection{Mass of the hosting star}
\label{subsub: uniform}
\begin{figure}[!ht]
\centering
\includegraphics[width=.43\textwidth]{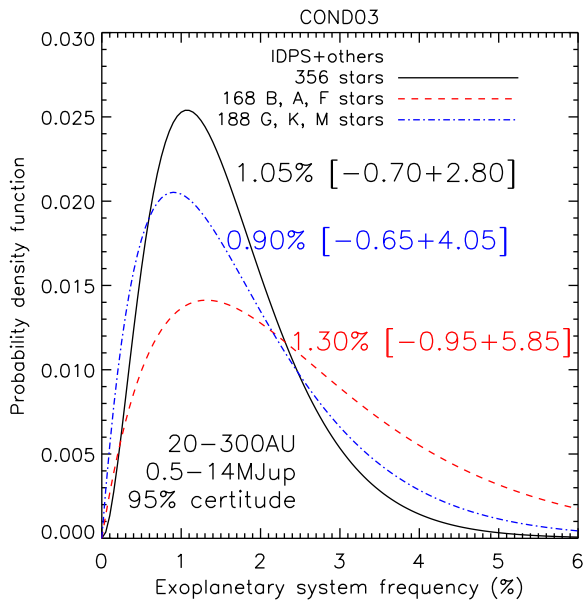}
\caption[]{\sl Probability density function of frequency of stars with at least one giant planet in the given intervals of masses and semimajor axes: $[m_{\mathrm{min}}, m_{\mathrm{max}}]=[0.5,14]\,$M$_J$ and $[a_{\mathrm{min}}, a_{\mathrm{max}}]=[20,300]$\,AU.}
\label{fig: plafreq}
\end{figure}

Under all the assumptions described in the previous sections, we derive the probability density function of the frequency of stars with at least one giant planet of mass in $[m_{\mathrm{min}}, m_{\mathrm{max}}]=[0.5,14]\,M_J$ and semimajor axis in $[a_{\mathrm{min}}, a_{\mathrm{max}}]=[20,300]$\,AU (black full line in Fig.\,\ref{fig: plafreq}).
The mode of the distribution (i.e., frequency for which the distribution is maximum) is $1.05\rlap{\textsuperscript{\tiny+2.80}}\textsubscript{\tiny-0.70}\ \%$. The error bars give a $95\%$ confidence interval. This means that $1.05\rlap{\textsuperscript{\tiny+2.80}}\textsubscript{\tiny-0.70}\ \%$ of stars harbors at least one giant planet of $0.5$-$14$\,M$_\mathrm{J}$ between $20$ and $300$\,AU with a $95\%$ certitude.

Surveying exoplanets with orbital periods shorter than four years around stars of 1 to 5 $M_\odot$ (i.e., semimajor axis of $<2$\,AU) and with masses up to~$30\,$M$_J$, \citet{reffert15} found that the frequency of stars with giant planets depends on the stellar mass, reaching a maximum for hosting stars of $\sim1.9 M_\odot$. We test this assertion on our sample for stars hosting at least one giant planet between $20$ and $300$\,AU and in the range of $0.5$ to~$14$\,M$_\mathrm{J}$. We split our sample into 168 massive stars (B, A, F types, $>1.1 M_\odot$, red dashed line) and 188 low-mass stars (G, K, M-types, $<1.1 M_\odot$, blue mixed line) and we overplot the associated probability density functions in Fig.\,\ref{fig: plafreq}. The mode of the distribution for the massive stars is $1.30\rlap{\textsuperscript{\tiny+5.85}}\textsubscript{\tiny-0.95}\ \%$ with $95\%$ confidence error bars. For the low-mass stars, we find $0.90\rlap{\textsuperscript{\tiny+4.05}}\textsubscript{\tiny-0.65}\ \%$. The two confidence intervals overlap and there is no significant difference at a $95\%$ confidence level between the frequencies of massive and low-mass stars that host an exoplanetary system. We cannot extend the \citet{reffert15} result to giant planets orbiting at large separations using our sample.

\subsubsection{Planet mass and semimajor axis distributions}
\label{subsub: powerlaw}
The various assumptions that are made to performed a statistical analysis impact the final results. Although studying each parameter effect is out of the scope of this paper, we tested one effect. Instead of assuming uniform distributions of planet masses and semimajor axes, we now use power laws similar to what was measured for close-in exoplanets by \citet{cumming08}. Hence, the number $\mathrm{d} N$ of simulated exoplanets whose mass is in the interval $[m,m+\mathrm{d} m]$ and semimajor axis is in the interval $[a,a+\mathrm{d} a]$ is proportional to $m^{-1.31}\,a^{-0.61}\,\mathrm{d} m\,\mathrm{d} a$. All the other assumptions being the same as before, we obtain that $2.30\rlap{\textsuperscript{\tiny+5.95}}\textsubscript{\tiny-1.55}\ \%$ of stars harbor at least one giant planet of $0.5$-$14$\,M$_\mathrm{J}$ between $20$ and $300$\,AU with a $95\%$ certitude. Thus, changing the assumption on the planet mass and semimajor axis distributions, the derived frequency increases by a factor of $\sim2$: the most probable frequency ranges from $1.05\,\%$ (uniform distributions, section~\ref{subsub: uniform}) to $2.30\,\%$, and the $95\,\%$ interval ranges from $[0.35\,\%, 3.85\,\%]$ (uniform distributions) to $[0.75\,\%,8.25\,\%]$. The two $95\,\%$ intervals overlap but the impact of the assumed planet mass and semimajor axis distributions is obvious in this case.

This result was expected because the power law distributions predict a lot more light and close-in planets than uniform distributions, which are planets the IDPS is less sensitive to. This leads to the open question: Which distribution shall we use for the planet mass and semimajor axis? On the one hand, we can argue that the distributions derived by \citet{cumming08} come from observations and should be a good a priori. On the other hand, we can wonder whether the formation process is or is not the same for close-in and outer giant planets, and thus, whether the distributions are or are not the same. In any case, future studies are needed to determine the impact of each assumption that is made to derive exoplanet frequencies. In the rest of the current paper, we provide the frequencies for both assumptions on the planet mass and semimajor axis.

\subsection{Comparision to other surveys}
\label{subsec: others}
We compare to previous results the frequencies of stars that harbor at least one giant planet as derived from our sample (Tab.\,\ref{tab: plafreq_comp}). We use mass and separation intervals as close as possible to those chosen by the other teams and we confirm all results. Having a larger number of stars and deep detection limits, we put stronger constraints reducing the estimated frequencies by a factor of two or more in almost all cases. Also, the difference between the frequencies derived for the two planet mass distributions that we consider (uniform or power laws) is small when we focus on planets more massive than $\sim1\,$M$_J$.
\begin{table*}[!ht]
\begin{center}
\renewcommand{\arraystretch}{1.2}
\begin{tabular}{lccccccl}
Stellar type&\multicolumn{3}{c}{Frequency (in $\%$)}&Separation&Planet Mass&Mass\&SMA&Reference\\
&&\multicolumn{2}{c}{$95\,\%$ certitude}&(AU)&(M$_J$)&distribution&\\
&Mode&Min.&Max.&&&&\\
\hline
BA&$1.90$&$0.50$&$10.05$&$59-460$&$4-14$&uniform law&This work\\
BA&$2.05$&$0.50$&$11.05$&$59-460$&$4-14$&power law&This work\\
BA&-&-&$20$&$59-460$&$>4$&&\citet{nielsen13}\textsuperscript{a}\\
\hline
BA&$2.70$&$0.65$&$14.30$&$25-940$&$4-14$&uniform law&This work\\
BA&$2.75$&$0.70$&$14.70$&$25-940$&$4-14$&power law&This work\\
BA&-&-&$50$&$25-940$&$>4$&&\citet{nielsen13}\textsuperscript{a}\\
\hline
FGK&$1.20$&$0.30$&$6.60$&$25-856$&$4-14$&uniform law&This work\\
FGK&$1.10$&$0.30$&$6.05$&$25-856$&$4-14$&power law&This work\\
F2-K7&-&-&$20$&$25-856$&$>4$&&\citet{nielsen10}\textsuperscript{a}\\
\hline
M&-&-&$8.30$&$9-207$&$4-14$&uniform law&This work\\
M&-&-&$8.60$&$9-207$&$4-14$&power law&This work\\
M0-M5&-&-&$20$&$9-207$&$>4$&&\citet{nielsen10}\textsuperscript{a}\\
\hline
M&-&-&$9.15$&$10-200$&$1-13$&uniform law&This work\\
M&-&-&$11.85$&$10-200$&$1-13$&power law&This work\\
M&-&-&$11$&$10-200$&$1-13$&&\citet{bowler15}\\
\hline
GKM&$1.30$&$0.35$&$7.00$&$30-900$&$5-14$&uniform law&This work\\
GKM&$1.05$&$0.30$&$5.85$&$30-900$&$5-14$&power law&This work\\
GKM&-&-&$2$&$\ge30$&$\ge5$&&\citet{kasper07}\textsuperscript{a}\\
\hline
FGKM&-&-&$3.30$&$10-500$&$0.5-13$&uniform law&This work\\
FGKM&-&-&$6.25$&$10-500$&$0.5-13$&power law&This work\\
FGKM&-&-&$57$&$10-500$&$0.5-13$&&\citet{lafreniere07a}\\
\hline
FGKM&-&-&$2.75$&$25-340$&$0.5-13$&uniform law&This work\\
FGKM&-&-&$5.20$&$25-340$&$0.5-13$&power law&This work\\
FGKM&-&-&$17$&$25-340$&$0.5-13$&&\citet{lafreniere07a}\\
\hline
FGKM&-&-&$2.40$&$50-230$&$0.5-13$&uniform law&This work\\
FGKM&-&-&$4.50$&$50-230$&$0.5-13$&power law&This work\\
FGKM&-&-&$10$&$50-230$&$0.5-13$&&\citet{lafreniere07a}\\
\hline
FGKM&$0.60$&$0.15$&$3.25$&$100-300$&$5-14$&uniform law&This work\\
FGKM&$0.60$&$0.15$&$3.30$&$100-300$&$5-14$&power law&This work\\
FGKM&-&-&$15$&$100-300$&$>5$&&\citet{chauvin15}\textsuperscript{a}\\
\hline
FGKM&$0.50$&$0.15$&$2.80$&$50-300$&$10-14$&uniform law&This work\\
FGKM&$0.50$&$0.15$&$2.75$&$50-300$&$10-14$&power law&This work\\
FGKM&-&-&$10$&$50-300$&$>10$&&\citet{chauvin15}\textsuperscript{a}\\
\hline
BAFGKM&$0.45$&$0.15$&$2.60$&$40-150$&$1-14$&uniform law&This work\\
BAFGKM&$0.75$&$0.20$&$4.10$&$40-150$&$1-14$&power law&This work\\
BAFGKM&-&-&$10$&$40-150$&$>1$&&\citet{chauvin10}\textsuperscript{a}\\
\hline
BAFGKM&$0.45$&$0.15$&$2.55$&$10-150$&$1-20$&uniform law&This work\\
BAFGKM&$0.85$&$0.25$&$4.70$&$10-150$&$1-20$&power law&This work\\
BAFGKM&-&-&$6$&$10-150$&$1-20$&&\citet{biller13}\\
\hline
BAFGKM&$0.35$&$0.10$&$1.95$&$10-100$&$5-70$&uniform law&This work\\
BAFGKM&$0.45$&$0.15$&$2.45$&$10-100$&$5-70$&power law&This work\\
BAFGKM&$1.70$&$0.52$&$4.90$&$10-100$&$5-70$&&\citet{brandt14}\\
\end{tabular}
\caption{\sl Frequencies of stars harboring at least one giant planet in given intervals of masses and semimajor axes derived from our work and from previous publications. Credible intervals are given for confidence levels of $95\%$. \textsuperscript{a}: the authors do not provide all the upper limits.}
\label{tab: plafreq_comp}
\end{center}
\end{table*}
Whereas each team makes different assumptions, all results are consistent with each other. This is interesting and future works should be done to understand which assumption(s) the derived frequencies mainly depend on.

\section{Conclusion}
We have completed the IDPS for giant planets around $292$ young nearby stars of all spectral types with a majority of massive stars. We developed a pipeline for a uniform processing of the data that have been recorded for $14$ years using several instruments: NIRC2/Keck II, NIRI/Gemini north, NICI/Gemini South, and NACO/VLT.  We achieved contrasts of $\sim12.5\pm2.5$ magnitude at $1''$ at H, CH4, K and Lp bands.

 We detected a total of $2,279$ point-like sources. Most of these sources were confirmed to be background objects. Four were confirmed to be exoplanets. They are the now well-characterized HR\,8799 planets \citep{marois08,marois10}. We also discovered $16$ stellar binary systems and $2$ triple stars.

We used the Bayesian formalism developed in \citet{lafreniere07a} to derive the frequency of stars with giant planets from the detection limits of the survey as well as the confirmed exoplanets. To complete the IDPS sample, we combined our results with the \citet{lafreniere07a} and \citet{chauvin10} surveys that mainly observed low-mass stars (G, K and M stars). The complete sample includes $356$ stars of all spectral types with a median age of $100$\,Myr and a median distance of $37$\,pc. We determined that $1.05\rlap{\textsuperscript{\tiny+2.80}}\textsubscript{\tiny-0.70}\ \%$ of stars harbor at least one giant planet of $0.5$-$14$\,M$_\mathrm{J}$ between $20$ and $300$\,AU. We also found no evidence that giant planets at large separations are more likely formed around BAF stars than around GKM stars. We confirm previous results reducing the measured frequencies of stars with at least one giant planet by a factor of two or more in almost all cases. The fact that the results of the different studies are consistent is encouraging, but we should keep in mind that each team uses different assumptions (planet atmosphere models, orbital parameter distributions, star aging, etc.) and one might wonder on which assumption(s) the derived frequencies mainly depend. For example, we showed in this paper that the assumption on the planet mass and semimajor axis distributions can change the conclusions and it will be essential to address this question to prepare the interpretation of the Gemini Planet Imager \citep{macintosh08} and the Spectro-Polarimetric High-contrast Exoplanet REsearch \citep{beuzit08} surveys. Most of their targets are part of the IDPS sample but the extreme adaptive optics systems will probe lighter and closer-to-their-star exoplanets, which will complete our knowledge of giant planets at large separations.

\section{Acknowledgment}

The authors acknowledge the referee for constructive comments and suggestions.

The authors wish to recognize and acknowledge the very significant cultural role and reverence that the summit of Mauna Kea has always had within the indigenous Hawaiian community. We are most fortunate to have the opportunity to conduct observations from this mountain.

This research has made use of the SIMBAD database, operated at CDS, Strasbourg, France \citep{wenger00}.

Some of the data presented herein were obtained at the W.M. Keck Observatory, which is operated as a scientific partnership among the California Institute of Technology, the University of California, and the National Aeronautics and Space Administration. The Observatory was made possible by the generous financial support of the W.M. Keck Foundation. 

This research has made use of the Keck Observatory Archive (KOA), which is operated by the W. M. Keck Observatory and the NASA Exoplanet Science Institute (NExScI), under contract with the National Aeronautics and Space Administration. 

Based on observations obtained at the Gemini Observatory (acquired through the Gemini Observatory Archive and the Gemini Science Archive), which is operated by the Association of Universities for Research in Astronomy, Inc., under a cooperative agreement with the NSF on behalf of the Gemini partnership: the National Science Foundation (United States), the National Research Council (Canada), CONICYT (Chile), Ministerio de Ciencia, Tecnología e Innovación Productiva (Argentina), and Ministério da Ciência, Tecnologia e Inovação (Brazil).

\bibliography{database.bib}   
\bibliographystyle{aa}   

\onecolumn

\input{1602_17_liststar.txt}

\input{1510_30_listobs_paper_newpa.txt}

\input{1510_30_contrast_archive_Tabpaper.txt}

\input{1510_30_bintab.txt}

\input{1510_30_cand_tab.txt}

\end{document}

%% file: 1602_17_liststar.txt
\begin{landscape}
\tiny\tt
\begin{longtable}{lllcccccccc}
\multicolumn{9}{c}{Table\,\ref{tab: targlist}}\\
Star&$\alpha$&$\delta$&Spec.&Dist.&\multicolumn{3}{c}{Age (Myr)}&Age method&Ref&IR\\
&(J2000.0)&(J2000.0)&type&(pc)&median&min&max&or Assoc.&&excess\\
\hline
\endhead
HR9&00:06:50.09&-23:06:27.14&F3V&39&10&5&30&BPMG&\citet{zuckerman04}&--\\
HIP682&00:08:25.75&+06:37:00.49&G2V&40&100&30&300&Li/X&&Y\textsuperscript{c}\\
2MASSJ00120761-1550327&00:12:07.61&-15:50:32.71&K7&95&100&30&300&Li/X/UV&&--\\
HIP1134&00:14:10.25&-07:11:56.84&F7V&46&30&20&50&Col&\citet{brandt14}&--\\
HD1160&00:15:57.30&+04:15:04.01&A0V&104&50&10&100&&\citet{nielsen12}&--\\
TYC1186-706-1&00:23:34.66&+20:14:28.75&K5&60&12&8&22&X/UV&&--\\
2MASSJ00281434-3227556&00:28:13.5&-32:28:01.49&F2&39&10&5&30&Li/Ha&&--\\
TYC596-145-1&00:29:18.60&+09:52:30.90&K5&182&10&5&30&UV&&--\\
\end{longtable}
\end{landscape}
\renewcommand{\baselinestretch}{1}

%% file: 1510_30_listobs_paper_newpa.txt
\begin{landscape}
\singlespace
\tiny\tt
\begin{longtable}{lcccccccc}
\multicolumn{9}{c}{Table\,\ref{tab: obslist}}\\
Star&Date&Nb of images$\times$integration(s)$\times$Coadd&FOV rotation&Telescope&Filter&Field lens&Mode&Other programs\\
&&&(deg)&&&Coronagraph&&\\
\hline
\endhead
GJ1002&2006-07-20&5$\times$0.2$\times$50&1.3&KE&Lp&-&V&-\\
HR9&2001-12-01&8$\times$15$\times$4&4.4&KE&Kp&w&PA&-\\
&2009-11-01&60$\times$20$\times$2&16.9&KE&Ks&600&V&-\\
&2009-12-05&45$\times$12$\times$5&22.3&GS&CH4-H4\%S&320&V&-\\
&2009-12-05&45$\times$12$\times$5&22.3&GS&CH4-H4\%L&320&V&-\\
HIP682&2004-07-13&8$\times$30$\times$2&17.0&KE&H&-&V&-\\
&2006-07-18&46$\times$30$\times$1&25.7&KE&Kp&-&V&-\\
&2012-09-04&78$\times$30$\times$1&41.3&KE&Ks&400&V&-\\
\end{longtable}
\end{landscape}

%% file: 1510_30_contrast_archive_Tabpaper.txt
\begin{landscape}
\singlespace
\tiny\tt
\begin{longtable}{lccccccccccccccccccc}
\multicolumn{20}{c}{Table\,\ref{tab: contrlist}}\\
Star&Date&Filter&0.3"&0.5"&1.0"&1.5"&2.0"&3.0"&4.0"&5.0"&6.0"&7.0"&8.0"&9.0"&10.0"&11.0"&12.0"&13.0"&14.0"\\
\hline
\endhead
HR9&2009-11-01&Ks&5.3&6.9&9.7&11.6&12.8&13.5&13.6&13.4&12.6&9.8&-&-&-&-&-&-&-\\
HR9&2009-12-05&CH4l&8.5&10.0&12.7&13.9&14.7&15.2&15.3&15.3&15.3&15.3&15.7&16.6&15.1&-&-&-&-\\
HR9&2009-12-05&CH4s&9.3&10.2&12.6&13.9&14.7&15.1&15.3&15.2&15.3&15.3&15.7&17.1&18.5&-&-&-&-\\
HIP682&2004-07-13&H&10.1&9.3&11.1&13.1&14.4&14.7&-&-&-&-&-&-&-&-&-&-&-\\
HIP682&2004-07-13&H&9.9&8.2&9.5&10.4&11.1&12.5&13.2&13.5&14.0&13.9&13.7&13.8&13.9&-&-&-&-\\
HIP682&2006-07-18&Kp&9.3&10.0&12.6&14.1&15.2&16.0&16.2&16.1&15.8&11.4&-&-&-&-&-&-&-\\
HIP682&2012-09-04&Ks&11.2&11.7&14.3&15.8&16.4&16.5&16.4&16.0&16.1&15.3&-&-&-&-&-&-&-\\
J00120761-1550327&2004-07-14&H&9.8&7.7&9.7&10.5&11.1&12.6&13.2&13.3&13.4&13.4&13.4&13.5&13.5&-&-&-&-\\
\end{longtable}
\end{landscape}

%% file: 1510_30_bintab.txt
\begin{landscape}
\singlespace
\begingroup
    \fontsize{8pt}{8pt}\selectfont
\begin{longtable}{lllcccccccc}
\multicolumn{11}{c}{Table\,\ref{tab: bin}}\\
Star&$\alpha$&$\delta$&$\rho$&$\theta$&Contrast&Filter&Date&Telescope&Status&Comments/References\\
&(J2000.0)&(J2000.0)&(arcsec)&(deg)&($\Delta$mag)&&&&&\\
\hline
\endhead
TYC1186-706-1&00:23:34.66&+20:14:28.75&1.74$\pm$0.01&136.7&0.6&Kcont&2008-12-17&KE&A\textsuperscript{a}&\\
&&&1.74$\pm$0.01&137.4&0.6&Kp&2009-07-30&KE&A\textsuperscript{a}&\\
TYC603-461-1&00:38:59.26&+08:28:41.28&0.10$\pm$0.01&54.14&0.04&H2v=1-0&2002-08-22&KE&A\textsuperscript{a}&\\
&&&0.06$\pm$0.02&6&0.1&CH4s&2006-07-17&KE&A\textsuperscript{a}&\\
HIP3589&00:45:50.89&+54:58:40.20&2.72$\pm$0.01&68.9&2.3&FeII&2004-07-14&KE&A\textsuperscript{b}&\citet{cvetkovic11}\\
1RXSJ004653.1+480844&00:46:53.20&+48:08:44.00&0.13$\pm$0.01&218.66&0.2&H2v=1-0&2002-08-22&KE&A\textsuperscript{a}&\\
&&&0.18$\pm$0.01&226&0.2&H2v=1-0&2005-07-27&KE&A\textsuperscript{a}&\\
&&&0.20$\pm$0.01&226&0.2&Kp&2006-09-03&KE&A\textsuperscript{a}&\\
\end{longtable}
\endgroup
\end{landscape}

%% file: 1510_30_cand_tab.txt
\begin{landscape}
\singlespace
\begingroup
    \fontsize{8pt}{8pt}\selectfont
\begin{longtable}{lcccccccccccccc}
\multicolumn{15}{c}{Table\,\ref{tab: cand}}\\
Star&$\alpha$&$\delta$&$\rho$&$\theta$&$\delta\rho=\rho\delta\theta$&$\Delta m$&Filter&$M$&$m_{\mathrm{min,max}}$&$a_\mathrm{proj}$&Date&Tel.&\#&Status\\
&(J2000.0)&(J2000.0)&(")&(deg)&(mas)&&&&($M_\mathrm{Jup}$)&(AU)&&&\\
\hline
\endhead
HR9&00:06:50.09&-23:06:27.14&19.691&337.347&40&--&Kp&--&--&770.1$\pm$13.0&2001-12-01&KE&1&?\\
HIP682&00:08:25.74&+06:37:00.48&5.472&268.037&14&15.47$\pm$0.49&Ks&--&--&--&2012-09-04&KE&1&B\\
&&&4.945&267.774&12&14.73$\pm$0.31&Kp&--&--&--&2006-07-18&KE&1&B\\
TYC596-145-1&00:29:18.60&+09:52:30.90&19.495&263.231&16&11.52$\pm$0.17&Kp&12.09$\pm$0.17&4-11&3666.3$\pm$669.6&2006-12-10&KE&1&?\\
HIP5186&01:06:22.82&+62:45:41.01&13.253&238.314&109&14.83$\pm$0.58&Kp&16.74$\pm$0.58&4-11&861.4$\pm$27.5&2013-08-23&GN&18&?\\
&&&11.939&55.381&82&--&K&--&--&776.0$\pm$24.3&2012-09-04&GN&1&?\\
&&&11.887&347.414&79&13.12$\pm$0.12&K&15.03$\pm$0.12&7-15&772.6$\pm$24.0&2012-09-14&GN&16&?\\
&&&11.885&55.423&77&12.82$\pm$0.10&K&14.73$\pm$0.10&7-17&772.5$\pm$23.9&2012-09-14&GN&1&?\\
\end{longtable}
\endgroup
\end{landscape}